\documentclass{aa}
\usepackage{graphicx}
\usepackage{epsfig}     
\usepackage{amsmath}
\usepackage{txfonts}
\usepackage{natbib}
\bibpunct{(}{)}{;}{a}{}{,}
\usepackage{url}
\begin{document}

  \title{Modelling the Galactic Interstellar Extinction Distribution 
\\ in Three Dimensions \thanks{The three dimensional results will be 
      available in electronic form
      at the CDS via anonymous ftp to cdsarc.u-strasbg.fr (130.79.128.5)
      or via http://cdsweb.u-strasbg.fr/cgi-bin/qcat?J/A+A/}   }
  \titlerunning{Modelling Interstellar Extinction}
  \author{D.J. Marshall
    \inst{1}, 
    A.C. Robin
    \inst{1},
    C. Reyl\'e
    \inst{1},
    M. Schultheis
    \inst{1}
    \and
    S. Picaud
    \inst{1}\inst{2}
  }
  \authorrunning{Marshall et al.}
  \offprints{D.J. Marshall}
  \institute{CNRS UMR6091,Observatoire de Besan\c{c}on, 
    BP 1615, 25010 Besan\c{c}on Cedex, FRANCE\\
    \email{marshall@obs-besancon.fr}
\and
    Instituto de Astronomia e Geof\'{i}sica (IAG/USP), S\~{a}o Paulo, Brazil
  }
\date{Received  / Accepted }


\newcommand{\bmin}{-10}
\newcommand{\bmax}{10}
\newcommand{\lmin}{-100}
\newcommand{\lmax}{100}
\newcommand{\step}{15\arcmin}

\newcommand{\barang}{$30\pm5\degr$ }
\newcommand{\barlen}{5.2 kpc }
\newcommand{\warpmaxpos}{460 pc }
\newcommand{\warpmaxneg}{460 pc }
\newcommand{\scheight}{$125^{+17}_{-7}$ pc }

\newcommand{\angpos}{$89\degr$ }
\newcommand{\radpos}{8.7 {\rm kpc} }
\newcommand{\slopepos}{0.14 }
\newcommand{\angneg}{$272\degr$ }
\newcommand{\radneg}{7.8 {\rm kpc} }
\newcommand{\slopeneg}{0.11 }

\newcommand{\changes}{}


\abstract
{}
{The Two Micron All Sky Survey, along with the Stellar Population
Synthesis Model of the Galaxy, developed in Besan\c{c}on, is used to
calculate the extinction distribution along different lines of
sight. By combining many lines of sight, the large scale distribution
of interstellar material can be deduced.}
{ The Galaxy model is used to provide the intrinsic colour of stars
and their probable distances, so that the near infrared colour excess,
and hence the extinction, may be calculated and its distance
evaluated.  Such a technique is dependent on the model used, however
we are able to show that moderate changes in the model parameters
result in insignificant changes in the predicted extinction.  }
{
This technique has now been applied to over 64000 lines of sight, each separated by \step, 
in the inner Galaxy ( $|l|\le$\lmax{\degr}, $|b|\le$\bmax{\degr} ). 
We have projected our three dimensional results onto a two dimensional plane in order to 
compare them with existing two dimensional extinction maps and CO surveys. {\changes 
We find that although differences exist due to
the different methods used or the medium traced, the same large scale structures are visible in each of the different maps.
Using our extinction map, we have derived the main characteristics of the large scale structure of the dust
distribution.
 The scale height of the interstellar matter is found to be \scheight. 
The dust distribution is found to be asymmetrically warped, in agreement with CO and 
HI observations of the ISM. However, the slope of the dust warp and the galactocentric distance where it starts are 
 found to be smaller than the values measured for the external HI disc:
 for positive longitudes
the angle is $\theta=$\angpos, it starts at \radpos from the Galactic center 
and grows with a slope of \slopepos, while at negative longitudes, the 
angle of the maximum is at $\theta=$\angneg, the starting radius \radneg and the 
slope \slopeneg.
Finally, the presence
of dust is detected in the Galactic bulge. It forms an elongated structure approximately \barlen long and lies
 at an angle of \barang
with respect to the Sun-Galactic centre direction.
 This may be interpreted as a dust lane along the
Galactic bar.}
This resulting extinction map 
will be useful for studies of the inner
Galaxy and its stellar populations. 
}
{}


\maketitle


\section{Introduction}
\label{sec:intro}
Interstellar extinction remains a serious obstacle for the observation of stars in the Milky Way, 
and for interpreting these observations in terms of Galactic structure. It has two main effects, that of dimming
sources and reddening them. Any stellar study near the Galactic plane
would need to correct for this effect to enable a
proper interpretation of the observations.  
Recent infrared surveys provide a superb tool 
to probe further into the densest parts of the Galaxy as the extinction suffered in the $K_{s}$ band
is approximately one tenth of that suffered in the visible.

Furthermore, extragalactic studies
often shun the so called ``Zone of Avoidance'' at around $|{ b}|\la$10{\degr} and towards the inner 
Galaxy
due to the amount of extinction and the high source confusion. 
{\changes The determination of interstellar extinction in this zone, which
covers about 10\% of the infrared sky, is thus an invaluable step to identifying low extinction windows at low 
Galactic latitude.}

Several studies have been undertaken to further our knowledge of the distribution of the extinction in our Galaxy. 
Most of them provide a two dimensional map of the extinction in different regions of the Galaxy in which they calculate
either the mean 
extinction or the total extinction for various lines of sight : {\changes \cite{froebrich2005} use accumulative star counts
from the 2MASS survey to construct a {\it relative} extinction map of the Galactic plane;
\cite{sumi2004} has created a two dimensional map of the {\it mean} extinction towards the Galactic bulge using red clump giants from the 
Optical Gravitational Lensing Experiment II; }
 \cite{dutra2003} and \cite{schultheis1999} 
determined the  {\it{mean}} extinction towards the same region using 2MASS data and DENIS data, respectively; 
\cite{schlegel1998} used the dust IR emission as observed by DIRBE to create an all sky two dimensional map of the {\it{total}} extinction
 integrated along each line of sight. However, according to \cite{arce1999}, the
\cite{schlegel1998} map overestimates the reddening by a factor of 1.3-1.5 in regions of 
smooth extinction with $A_{V}>0.5$, and underestimates it in regions of steep extinction gradients. \cite{cambresy2005}
 also noticed that the correlation of the near-infrared extinction to the 
far-infrared optical depth shows a discrepancy of about 30\% for $A_V>$1 mag, probably due to 
the ubiquitous presence of fluffy grains.

Other authors have determined the distribution of extinction
 along different lines of sight using various methods: \cite{fitzgerald1968} 
provided the local distribution of extinction up to a few kiloparsecs using photoelectric photometry. Later on, 
\cite{neckel1980} measured extinction and distances towards individual O,B and A stars. 
These studies provide limited  resolution for the extinction distribution due to the scarceness of the 
distribution of hot stars, often related to clouds. They do not provide the distribution of the diffuse extinction. 
The study of \citet{berdnikov1990} is limited to a sample of 615 stars in 9 Kapteyn Selected Areas and shows 
considerable dispersion in extinction among the stars in a given direction. \cite{arenou1992} mapped the local
 extinction from a collection of spectral and photometric data. The validity of this map is limited to a 
distance of about 1 kpc, hence it cannot be used at $|b|< $10{\degr}, where a significant part of the extinction 
is situated at distances larger than 1 kpc. In addition, 
the resolution of the map is of the order of several degrees. 
\cite{cambresy1999} determined detailed maps of extinction in nearby Giant Molecular Clouds from USNO star
 counts, which are very useful for studies of these clouds, but not outside. He also considered the whole sky extinction
 from USNO star counts, but still with a two dimensional approach, and with a limited spatial resolution of 6{\degr}
\citep{cambresy1999sf}. A great deal of other determinations can also be found in the literature
 which concern one or a few lines of sight. Putting all these maps together leads to inhomogeneous maps with 
poorly defined accuracy. They also have the problem of matching different results at the boundaries between
authors \citep{hakkila1997}.

The relationship between infrared colour excess and extinction is well documented, and has been
in use for some time \citep{jones1984, lada1994}. In this type of study, stars 
that are reddened due to their location behind a cloud
are compared to nearby, supposedly un-reddened, stars. Our approach is to compare observed 
infrared colours with those from the Besan\c{c}on model of the Galaxy.
Using this technique, with which we are able to calculate
the extinction along the line of sight, we have created a map of $K_{s}$ band extinction ($A_{Ks}$)
 in the plane of the inner Galaxy 
($|l|<$100{\degr}, $|b|<$15{\degr}) and in three dimensions. Our method for 
determining extinction does not assume any 
model for the dust distribution. Instead, we compare the difference in the \mbox{$J{-}K_{s}$} 
colour distribution of stars from
the Besan\c{c}on model of the Galaxy \citep{robin2003}  to those from 2MASS observations \citep{cutri2003}; 
the difference between the two is then attributed
to interstellar extinction. The $J{-}K_{s}$ colour index provides an excellent probe of interstellar extinction,
 in the Galactic plane where it is high,
as radiation at two microns is less absorbed than visible light by around an order of magnitude.
This technique can therefore provide a global picture
of the three dimensional distribution of interstellar dust. 

Our use of the 2MASS data is discussed in \S\ref{sec:data}, the generation of catalogues of  stars
using the Besan\c{c}on model of the Galaxy in \S\ref{sec:besac}. 
In \S\ref{sec:method} we describe how we transform a difference
in colour between observed and modelled stars into a prediction of the three dimensional extinction along a line of sight.
The three dimensional results are presented in \S\ref{sec:results}; these are compared to 
observations of CO and to other
extinction data, and the large scale characteristics of its distribution are discussed. 
The sensitivity of the results to changes in model parameters is estimated and
limits and possible bias are discussed in \S\ref{sec:discussion}.
 Our results are summarised in \S\ref{sec:conclusion}.


\section{2MASS Point Source Catalogue}
\label{sec:data}
\subsection{Description of the survey}
The Two Micron All Sky Survey (2MASS) is a ground based survey which
uniformly scanned the entire sky in three near-infrared bands ($J$, $H$ \& $K_{s}$). 
Amongst its final products is the point source catalogue (PSC) which includes 
point sources brighter than about 1 mJy in each band, 
with signal-to-noise ratio (SNR) greater than 10, using a pixel size of 2.0\arcsec.
Each star has accurate photometry and astrometry as defined in the level 1 
requirements \citep{cutri2003}.

The quality of the data was destined to meet or surpass all of the level 1 requirements.
We list three of them below : 
\begin{itemize}
\item 
Photometric sensitivity of 10-$\sigma$ at 15.8, 15.1, 14.3 mag at $J$, $H$, $K_{s}$ respectively for $|b|>$10{\degr}
\item 
Completeness $>$0.99 at 10-$\sigma$ sensitivity limits
\item
Reliability $>$0.9995
\end{itemize}

These requirements were met for the entire {\it unconfused} sky. 
However, the 2MASS survey has no level 1 requirements for the Galactic Plane - these specifications apply only to high
Galactic latitudes.
Within approximately a 6{\degr} radius of the Galactic centre,
the average magnitude for a
SNR=10 source is up to $\sim$2 mag brighter than the typical SNR=10 source in unconfused regions. 
This is also seen 
for various regions 
in the Galactic plane at $|b| \la$ 3{\degr} and $\pm$  60{\degr} from
the Galactic centre. 

As well as containing default magnitudes and associated uncertainty for each star, the 2MASS PSC
also contains a number of flags to aid the user in selecting stars appropriate for the study. We make
use of the Read Flag (rd\_flg), which indicates the method used to determine the default magnitude of the source,
 and the Photometry Quality Flag (ph\_flg) which gives an indication of the quality
of the photometric determination {\changes (signal to noise ratio, measurement error)}. More information on the 
different flags and their meanings can be found in \cite{cutri2003}.

{\changes The final 2MASS project also includes an extended source catalogue (XSC). 
By using a time varying PSF as well as
the  $J{-}K_s$ and $H-K_s$ colour of the sources, 
 \cite{cutri2003} were able to separate the extended sources from the point sources.
After this classification, the final 2MASS PSC contains only point sources with a reliability of over 98\%. 
}

\subsection{Completeness of the 2MASS PSC}
\label{sec:2masscomp}

We are interested in regions in the Galactic Plane, where the completeness 
varies as a function of stellar density and can be as much as two magnitudes brighter
than high latitude fields. 
The PSC can be compared quantitatively with model simulations only where the former
is shown to be complete. As such, we must 
compute the completeness limit field by field.

The 2MASS definition of completeness is 
"the faintest magnitude bin which recovers 99\% of the expected source counts". However, 
for their atlas images, \cite{cutri2003} suggest another method, described below, which we have adopted. 
Star count histograms are constructed for each field at  
0.2 magnitude intervals for point sources with magnitudes between 9 and 18 mag and 
{\changes where a reliable estimate of the photometric error could be determined in the appropriate band.
This latter condition  excludes only a few percent of the total number of stars in a field except in high
confusion fields, for example near the bulge or in the Galactic plane, 
where as many as 20\% of the stars in the 2MASS PSC
may not have an associated photometric error.
}  
The completeness in a particular field is defined to be the bin
that contains the highest star count. {\changes Stars detected by 2MASS but with photometry so poor that
the error could not be determined are not used for the calculation of the limiting magnitude. This ensures
that we do not overestimate the completeness of the observed stars which would introduce a large
number of modelled stars with no observational counterpart.}

\subsection{2MASS selection criteria}

As stated, there are a number  of  flags for each star in the PSC in order to select
them based on the quality of their photometry, or to reject them due to 
the risk that they are contaminated by nearby bright objects. 
However, if we are too severe
with our selection we will reject valid stellar detections. We will be comparing the selected 
stars with the Besan\c{c}on model; 
it thus follows that 
we must select 
all point sources from the PSC that correspond to actual stellar detections.

As we are interested in using the difference in observed and modelled $J{-}K_{s}$ colour, we require detections in these
two bands in order to admit a star into our calculation. However, a point source
 that has been detected in the $J$ and $K_{s}$ bands 
without a corresponding detection in the $H$ band is not likely a valid stellar detection.
Therefore, we restrict our selection to stars that have been
detected in all three bands (2MASS rd\_flag $\neq$ 0). {\changes Stars above the completeness limit in either the 
$J$, $H$ or $K_{s}$ bands are also rejected. 
We use the $J-K_s$ colour index as it is less sensitive to metallicity and
gravity than the $H-K_s$ or $J-H$ colour indices.}

This ``full'' catalogue includes all detected stars in the particular 2MASS field, even those with 
less accurate photometry. 
However, as they are real detections they must be included in our comparison with the Besan\c{c}on model.

\subsection{Missing fields in the PSC}

The 2MASS catalogue contains a small number of ``physical'' or ``effective'' gaps, 
due either to unscanned areas of the sky or the removal of valid sources near tile
boundaries, respectively. The total area of these gaps in our results amounts to less than 0.04 square degrees.

Other gaps in the 2MASS catalogue, which add up to a larger area in our results,
 include regions where bright stars have been removed, 
effectively creating a hole in the stellar density. 
All stars within the radius of the removed star are then excluded from the field of view;
as these ``missing'' stars sample all stellar populations and distances, 
they do not introduce any bias or error into our extinction determination.

\section{The Galactic model}
\label{sec:besac}
\subsection{Description of the model}
The stellar population synthesis model of the Galaxy constructed in Besan\c{c}on \citep{robin2003},
hereafter called the Galactic model,
  is able to simulate
the stellar content of the Galaxy by modelling four distinct stellar populations:
the thin disc, the thick disc, the outer bulge and the spheroid. It also takes
into account the 
dark halo and a diffuse component of the interstellar medium. It can be used to generate 
stellar catalogues for any given direction, and 
returns information on each star such as
magnitude, colour, and distance as well as kinematics and other stellar parameters.

The approach of the Galactic model is semi-empirical as it is based on 
theoretical grounds (for example stellar evolution, galactic evolution and galactic dynamics) and is constrained
 by empirical observations (the local luminosity function for example). 
The Galactic potential is calculated in order
to self-consistently constrain the disc scale height. In addition, the model
includes observational errors and Poisson noise to make it ideal for 
direct comparison with observations.

The Galactic model has been developed 
to return results in the near-infrared and visible filters.
It has been shown to reproduce the stellar content in the infrared to a 
high degree of accuracy (see \S\ref{sec:change_params}), 
and is a powerful tool to extract the extinction 
information embedded in the 2MASS observations.

\subsection{Model parameters}
In this study, limited to the inner Galaxy, the bulge and
thin disc are the dominant populations. Hence, the thick
disc and spheroid model parameters will not be described here,
but can be found in \cite{robin2003}.

\subsubsection{Thin disc}
The thin disc is described by an evolutionary scheme with
a two-slopes initial mass function and a constant star
formation rate over the past 10 Gyr.
The thin disc is divided into 7 age components, the first of which is called the young disc (age $<0.15$ Gyr).
The other six components define the old disc.

The old thin disc density distribution is modelled using the \citet{einasto1979} law. The distribution of 
each old disc component is described by an axisymmetric ellipsoid with an axis ratio depending 
on the age; the density law is described by the subtraction of two ellipsoids :

\begin{equation}
\rho_{\rm d}=\rho_{\rm d0} \times \left[ {\rm disc} - {\rm hole} \right]
\label{eqn:params}
\end{equation}
where
\begin{eqnarray*}
& & {{\rm disc}= exp{\left(-\sqrt{0.25+\left(\frac{R}{R_{\rm d}}\right)^2 + 
\left(\frac{Z}{\epsilon \times R_{\rm d}}\right)^2 }\right)}} \\
& & {{\rm hole}=exp{\left(-\sqrt{0.25+\left(\frac{R}{R_{\rm h}}\right)^2+
\left(\frac{Z}{\epsilon \times R_{\rm d}}\right)^2}\right)}} 
\end{eqnarray*}
and :
\begin{itemize}
\item
$R$ and $Z$ are the cylindrical galactocentric coordinates
\item
$\epsilon$  is the axis ratio of the ellipsoid. Recently
revised  axis ratios of the 6 age components
of the old thin disc can be found in \cite{robin2003};
\item
$R_{\rm d}$ is the scale length of the disc and is 2.510 kpc
\item
$R_{\rm h}$ is the scale length of the hole and is 680 pc 
\item
the normalization $\rho_{d0}$ is deduced from the local luminosity
function \citep{jahreiss1997}, assuming
that the Sun is located at R  = 8.5 kpc and Z  = 15 pc.
\end{itemize}

The two values for $R_{\rm d}$  and $R_{\rm h}$ are recent results from the method described in \cite{picaud2004}. 
The equation of the old disc differs from that found in \cite{robin2003};  the version presented here 
improves the 
modelling of the disc hole.

The young thin disc density distribution is :
\begin{equation}
\label{eqn:params_young}
\rho_d=\rho_{d0} \times \left[ exp\left(-\left(\frac{a}{R_d}\right)^2\right) 
- exp\left(-\left(\frac{a}{R_h}\right)^2\right) \right]
\end{equation}
{\changes where $a^2=R^2+\left(\frac{Z}{\epsilon}\right)^2$, $R$, $Z$ and $\epsilon$ have the same 
significance as above and the scale lengths for the young disc are $R_d=5$ kpc and $R_h=3$ kpc.}

The stars from the young disc 
represent approximately 0.5\% of the stars in this study, thus the impact of this population 
in our method
is negligible.

\subsubsection{Outer Galactic bulge}
\label{sec:outerbulge}
An analysis of DENIS data in the region with Galactic coordinates
$-8{\degr} < l <12{\degr}$ and
 $|b|<4{\degr}$ allowed \cite{picaud2004} to constrain the shape of the triaxial
bulge, as well as its age. The bulge is found to be boxy,
prolate, with axis ratios 1 : 0.30 : 0.25,  and oriented $10.6 \pm 3{\degr}$ with respect to the Sun-centre
direction. A full description of the parameter values of
the bulge density law can be found in \cite{picaud2004}.

\subsubsection{Spiral structure and Galactic warp}
\label{sec:spiral}
The version of the Galactic model we use here does not include any spiral structure.
The stars that we will be using for our method, mostly old K\&M giants, are, according to the
Galactic model, well over 1 Gyr old which corresponds to about 4 Galactic rotations.
The effect of the spiral structure is thus assumed to be negligible for stars older
than 1 Gyr.

The stellar warp is incorporated 
into the Galactic model as described
in \cite{robin2003}. Here we give just
the relevant values.
{\changes The Galactic warp is modelled as a shift 
of the galactocentric coordinates perpendicular
to the
plane by a linear factor equal to 0.18 beyond a 
galactocentric distance of 8.4 kpc \citep{derriere2001a,derriere2001b}}. This effect is maximum
for an angle of 90{\degr} from the Sun Galactic centre direction.

\subsection{Diffuse extinction}
\label{sec:model_ext}
At the time of writing, the Galactic model available online (\url{http://www.obs-besancon.fr/modele/modele.html})
 is able to simulate interstellar extinction as
diffuse extinction and as individual clouds. For the former, it is approximated using a double
exponential disc with an {\it ad hoc} local normalisation (usually 0.7 mag kpc$^{-1}$ for high latitude lines
of sight). For the latter, any number of clouds can be added along the line of sight, defined by their distance and 
extinction ($A_{V}$). A cloud covers the entire field of view and therefore 
affects {\it all} the stars at distances greater than it.

\subsection{Photometric errors in the model}
\begin{figure}
  \begin{center}
    \leavevmode
	{\epsfig{file=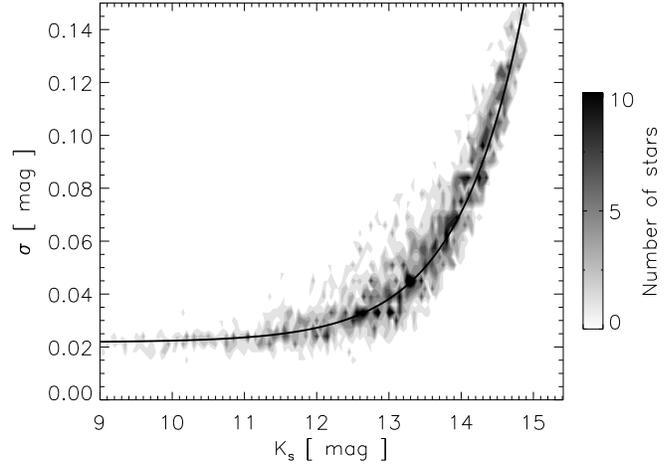,width=\linewidth}}
   \end{center}
  \caption{Grey-scale representation of the photometric error as a function of the $K_{s}$ 
magnitude for the 2MASS observations in the field $(l,b)=(50\degr,0\degr)$. 
    The solid line shows our best fit using Eq.\ref{eqn:photerr}}
  \label{fig:photerr}
\end{figure}
The 2MASS PSC supplies photometric errors for each band and for each star, where the 
conditions (atmospheric, crowding) permitted their determination. 
The Galactic model is able to simulate different photometric error laws
to emulate the response of different detection methods (CCD, photographic plates, etc).
Photometric errors are
included on each photometric band independently, the error
being a function of the magnitude (as usually observed). 
An exponential function is able to 
reproduce typical CCD photometric errors very well.

The error law,
as implemented in the model, is :
\begin{equation}
\label{eqn:photerr}
\sigma({m})={A}+  {\rm e}^{({C} \times m - {B})}
\end{equation}
where $m$ is the magnitude in the band considered, and $A$, $B$ and $C$
are fitted to the set of data with which the model predictions are
going to be compared.  {\changes Fig.\ref{fig:photerr} shows the 
2MASS photometric errors for the $K_s$ band in the field ($l,b$)=(50\degr, 0\degr). The solid line is the best fit
using Eq.\ref{eqn:photerr}.}

{\changes The photometric errors in the 2MASS PSC are not constant across
the entire sky, as they depend on local atmospheric conditions and stellar density.} 
To perform the fit {\changes for a given field}, we select the 
2MASS stars {\changes in that field where a reliable estimate of the photometric error could be determined in the appropriate band.}
The parameters of Eq. \ref{eqn:photerr} are 
then determined by fitting the error law to the selected 2MASS stars. {\changes This enables us to define
an average error with respect to apparent magnitude, field by field, for the stars in the 2MASS PSC}.

\subsection{Limiting magnitudes in the model}

In order to closely model the 2MASS stars, we cut the modelled stars at the 2MASS completeness and 
at a $J$, $H$ and $K_{s}$ magnitude of 9 as a lower limit. The faint limit ranges from 
m$_{Ks} \sim$10.5 to  $\sim$14.4, and $m_{J} \sim$12.0 to $\sim$15.8, depending on the
crowding of the field. 
{ \changes
As the magnitude increases, the reliability of the 2MASS catalogue in the {\it Galactic plane} decreases. 
Therefore, to ensure the we have the highest reliability 
we fix the faint magnitude limit of the $K_s$ band to a maximum of 12. The 
maximum magnitude used for the $J$ band, however, is chosen to be the completeness limit of the 2MASS observations
in order to detect sources with high extinction.
}


\section{Extinction from 2MASS and the Galaxy model}
\label{sec:method}

{\changes Interstellar dust absorbs and scatters electromagnetic radiation, the effect
being stronger for radiation at shorter wavelengths. This interstellar extinction 
results in observed stars appearing redder than they would in the absence of dust. 
  
The Galactic model provides us with simulated, initially unreddened stars and the 2MASS PSC
contains observed, reddened stars. If we assume that the Galactic model successfully predicts the 
average intrinsic colour and distance of stars in the Milky Way, then we may suppose that the difference between
the modelled and observed stellar colour distributions is due solely to the effects of extinction. 

The method that we detail in this section enables us to determine the extinction, as a function of distance and
along any particular line of sight, which minimises 
the difference between the modelled and observed stellar colour distributions.

}
\subsection{Colour distance relation}
\label{sec:colourdist}
\begin{figure}[t]
  \begin{center}
    \leavevmode
   \centerline{\epsfig{file=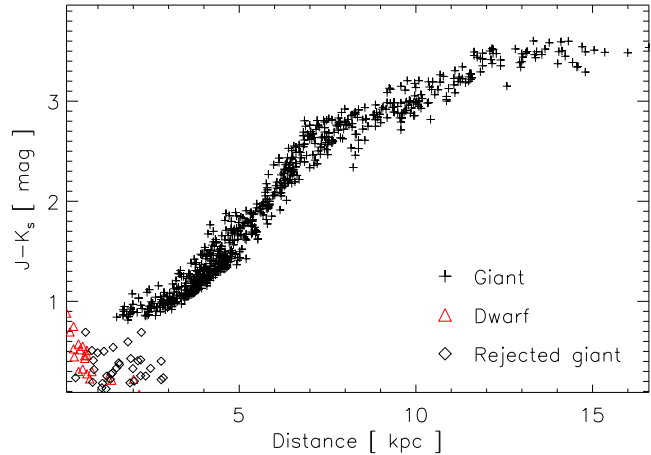,width=\linewidth}}
   \end{center}
  \caption{
Simulated colour-distance distribution towards l=330\degr b={1.5\degr}. 
Crosses are giants, triangles are dwarf stars and the diamonds denote rejected giants (\S\ref{sec:dwarfs}).
The group of stars in the lower left corner is dominated by dwarfs and sub-giants.
There is a relation between colour and distance for the
giants, first dominated by K\&M stars, then, at larger distances by RGB/AGB stars.}
  \label{fig:colour_dist}
\end{figure}
{\changes 

In Fig.\ref{fig:colour_dist} we present an example field from the Galactic model 
towards l=330\degr, b=1.5\degr;
the $J{-}K_{s}$ distribution is clearly a function of distance. 
The dwarf and sub-giant populations 
 are conspicuous
on the bottom left of Fig. \ref{fig:colour_dist} as they do not follow the same colour 
distance relation as the regular giant population.
Dwarf stars are observed locally; they are too faint to be seen at large distances. 
As such they do not suffer much interstellar reddening, so
the reddest (faintest) dwarfs detected will, contrary to the giants, be the closest.
These dwarf stars are removed from our stellar selection as described in \S\ref{sec:dwarfs}. Most
of the sub-giants which are not on the same sequence as the giants are also removed in this process.

The colour distance relation for the giants holds due to the fact that 
in the Galactic Plane the stars that dominate the counts at {9 $< K_{{s}}<$14}  
are mostly K0 to K2 giants and RGB/AGB giants; the latter dominate at 
larger distances due to their high luminosities.
Stars which are homogeneous in intrinsic $J{-}K_{s}$ colour and in 
absolute  magnitude $M_{K}$  would  have the same colour in a region with no extinction, 
showing a narrow histogram of colour.
As extinction is an increasing function of distance, 
the  $J{-}K_{s}$ colour of these 
stars will naturally increase with distance.

In the following we show how we compute the extinction, and its distribution in distance along the line of sight,
for a particular field. 
We use the Galactic model to simulate 
a variable proportion of K and M giants, as well as RGB/AGB stars, as a function of K 
magnitude along each line of sight, resulting in a realistic $J{-}K_{s}$ distribution.
In this way, the colour distance property of giants can be used to  translate 
the shape of a $J{-}K_{s}$ histogram of 2MASS stars in a given direction 
 into a distribution of the extinction along the line of sight.

}

\subsection{NIR colour excess}
\label{sec:nir_colour}

{\changes As mentioned above, interstellar dust has the effect of  reddening starlight.
Thus a difference in stellar colour (observed - intrinsic) can be  used to deduce the amount of extinction
the starlight has suffered.
For a group of stars assumed to be at a similar distance, the  difference between their mean intrinsic colour and
their mean observed colour (the mean colour excess) yields
the total mean extinction at that distance. Using a reddening law  \citep[for example][]{mathis1990}, we may
write:
\begin{equation}
\label{eqn:Ak_basic}
\overline{A_{{Ks}}} = 0.67 \times \overline{E(J{-}K_{s})}
\end{equation}
where $A_{{Ks}}$ is the extinction in the band $K_{s}$, $E(J{-}K_{s}$) is the $J{-}K_{s}$ colour excess
and the bar denotes that we have taken the mean for the group of stars.

Instead of using the intrinsic colour of the stars as reference, we use
simulated stellar colours from the Galactic model and we compute
the {\it difference} in extinction between observed and modelled stars:
\begin{equation}
\label{eqn:dAk}
\delta A_{{Ks}} = 0.67 \times \left[(\overline{J{-}K_{s}})_{\rm  obs} -  (\overline{J{-}K_{s}})_{\rm sim}\right]
\end{equation}
where the obs and sim subscripts denote observed and simulated  colours, respectively.
In other words, with Eq.\ref{eqn:dAk} we are calculating the  extinction necessary
to reduce, as much as possible, the difference in colour between  simulated and observed colour histograms.
This equation may be used if the simulated stars have been reddened  or not, as it simply translates
the difference in colour between simulated and observed stars into a  difference in extinction.
}

{\changes In order to calculate the
extinction distribution as a function of distance
for a particular line of sight and field size, we proceed as follows.
We start by applying a simple extinction distribution, as explained  in \S\ref{sec:model_ext}, to the simulated stars.
The local normalisation is chosen in order to minimise the difference  in the $J{-}K_{s}$ colour distributions of the
observed and simulated stars.
The simulated, reddened stars are then cut at the faint magnitude limits
for the field of the 2MASS observations. The dwarf stars, identified  as explained below
in \S\ref{sec:dwarfs}, are removed from both the observations and the  model.

Next, both the simulated and observed stars are sorted by increasing
$J{-}K_{s}$ colour;
due to the colour distance relation
mentioned above, this means that we are effectively sorting the stars  by increasing distance.
We then bin the simulated stars by colour;
 the number of  stars in a bin is chosen
such that the median distance of the simulated stars
in each successive bin increases.

The observed stars are put into the same number of bins as the  simulated stars, each bin containing the same
relative number of stars:
\begin{equation}
  n_{{\rm obs}_i} = \frac{N_{\rm obs}}{N_{\rm mod}} n_{{\rm mod}_i}
\end{equation}
where $N_{\rm obs}$ and $N_{\rm mod}$ are the total number of stars in the observations and model, respectively, and $n_{{\rm obs}_i}$ and $n_{{\rm mod}_i}$ denote the number of stars in a particular bin for the observations and the
Galactic model, respectively. 
We may then
obtain the magnitude of the extinction by applying Eq.\ref{eqn:dAk},  bin by bin, and assuming that the
extinction calculated is at the median distance of the simulated  stars in that bin. Using the relative number of
stars makes the method less sensitive to a difference in the number of stars between the Galactic model and
the observations, hence to the assumed large scale structure of the
stellar populations (\S\ref{sec:change_params}).
} 

\begin{figure}[t]
  \begin{center}
  \leavevmode
{\epsfig{file=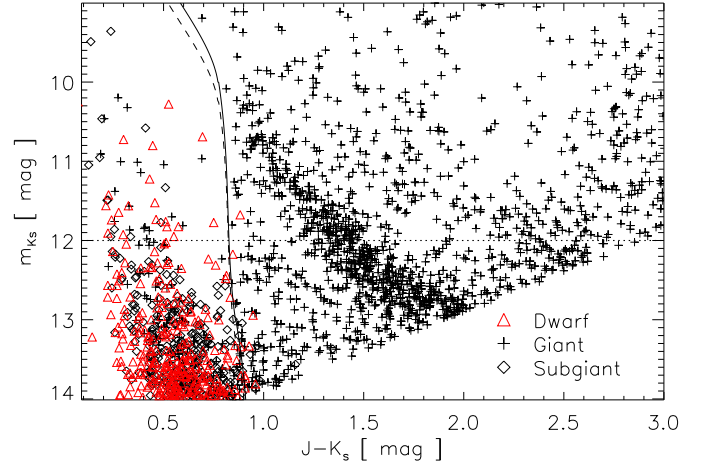,width=\linewidth}} 
  \caption{Dwarf star removal for the field ($l,b$)=(330,1.5). 
{\changes The triangles are dwarf stars, the diamonds are sub-giants and the crosses are regular giants.
 The lines running from top to bottom near $J{-}K_s=0.9$ are the cut-off for dwarf stars before (dashed)
and after(solid) the first determination of the extinction. The dotted line at $m_{Ks}=12$ shows our faint magnitude cutoff.
}}
  \label{figure_dwarfs}
\end{center}
\end{figure}
\subsection{Removing the dwarf stars}
\label{sec:dwarfs}

In Fig. \ref{figure_dwarfs} the colour magnitude diagram 
of the Galactic model for the line of sight 
$(l,b)=(330,1.5)$ is shown. 
The giant and dwarf populations occupy distinct regions in the diagram: 
the dwarf population is on the left hand side, the giants are on the right. 
The line running down the middle
indicates the upper limit of dwarf $J{-}K_{s}$ colour as a function of magnitude (described below). 
We select all stars redder than this upper limit; what will be left will be almost exclusively K \& M  giants {\changes and
RGB/AGB giants}. 

{\changes For a given apparent magnitude, we are able to define a maximum $J{-}K_{s}$ colour for the dwarf stars as 
follows. 
If we suppose that, on average, the closest stars 
are at 100 pc, then we can assign this distance to the reddest dwarf stars as these will be the closest.  
The absolute magnitude of these dwarf stars at a given apparent magnitude can then be calculated, initially
assuming that they do not suffer any extinction.
Using the main sequence relation, we are able to convert this absolute magnitude to a $J{-}K_{s}$ colour index. This
colour index thus defines the upper limit for the dwarf stars.}
However the dwarf stars do suffer some extinction; after an initial estimate of the extinction has been obtained
using the giants (\S\ref{sec:nir_colour}), we
 redden the upper limit used according to the extinction found at the distance of the dwarf stars.
This gives us a new selection of giants with which we may recalculate the extinction. {\changes Fig.\ref{figure_dwarfs} shows
the $J-K_s$ cut-off before (dashed line) and after (solid line) 
the first extinction determination. As can be seen, this is not a large shift 
as the dwarf stars are indeed local.}

\subsection{Iterative method}
\label{sec:chi2}
After determining the extinction distribution as
described above we apply this extinction to the stars from the Galactic model for the 
particular line of sight. To test the resulting distribution we construct a $J{-}K_{s}$ histogram
of the adjusted model  and the 2MASS observations (including the dwarf stars) - these can be compared by means of a 
$\chi^2$ test.

We compute the $\chi^2$ statistic 
for the two binned sets of data as detailed in \citet{press1992} :
\begin{equation}
\label{eqn:X2}
\chi^2 = \sum_i \frac{(\sqrt{N_{\rm obs}/N_{\rm mod}} n_{{\rm mod}_i} - 
\sqrt{N_{\rm mod}/N_{\rm obs}} n_{{\rm obs}_i})^2}{n_{{\rm mod}_i}+n_{{\rm obs}_i}}
\end{equation} 
where $n_{{\rm obs}_i}$ ($N_{\rm obs}$) and $n_{{\rm mod}_i}$ ($N_{\rm mod}$) 
are the number of stars in the $i^{\rm{th}}$ bin of the $J{-}K_{s}$ histogram 
(total number of stars along the line of sight) of the observations and model, respectively.  
The term involving the total number of stars ensures that 
the $\chi^2$ statistic will be lowest when the shapes of the two 
histograms are the same, regardless of any difference in the total number of stars.

{\changes
After having adjusted the extinction} some stars will be beyond the completeness limit for the field
 whereas others, intrinsically bright, will be
dim enough to be admitted into our selection. We therefore repeat the process using the adjusted stars and
recompute the $\chi^2$ statistic. We continue
this iterative process until we find a minimum in the $\chi^2$ statistic; the line of sight extinction giving rise
to this value is then taken to be the best fit for the particular line of sight. 
{\changes This is 
usually accomplished in 
under 10 iterations.}

\begin{figure}
  \begin{center}
  \leavevmode
      {\epsfig{file=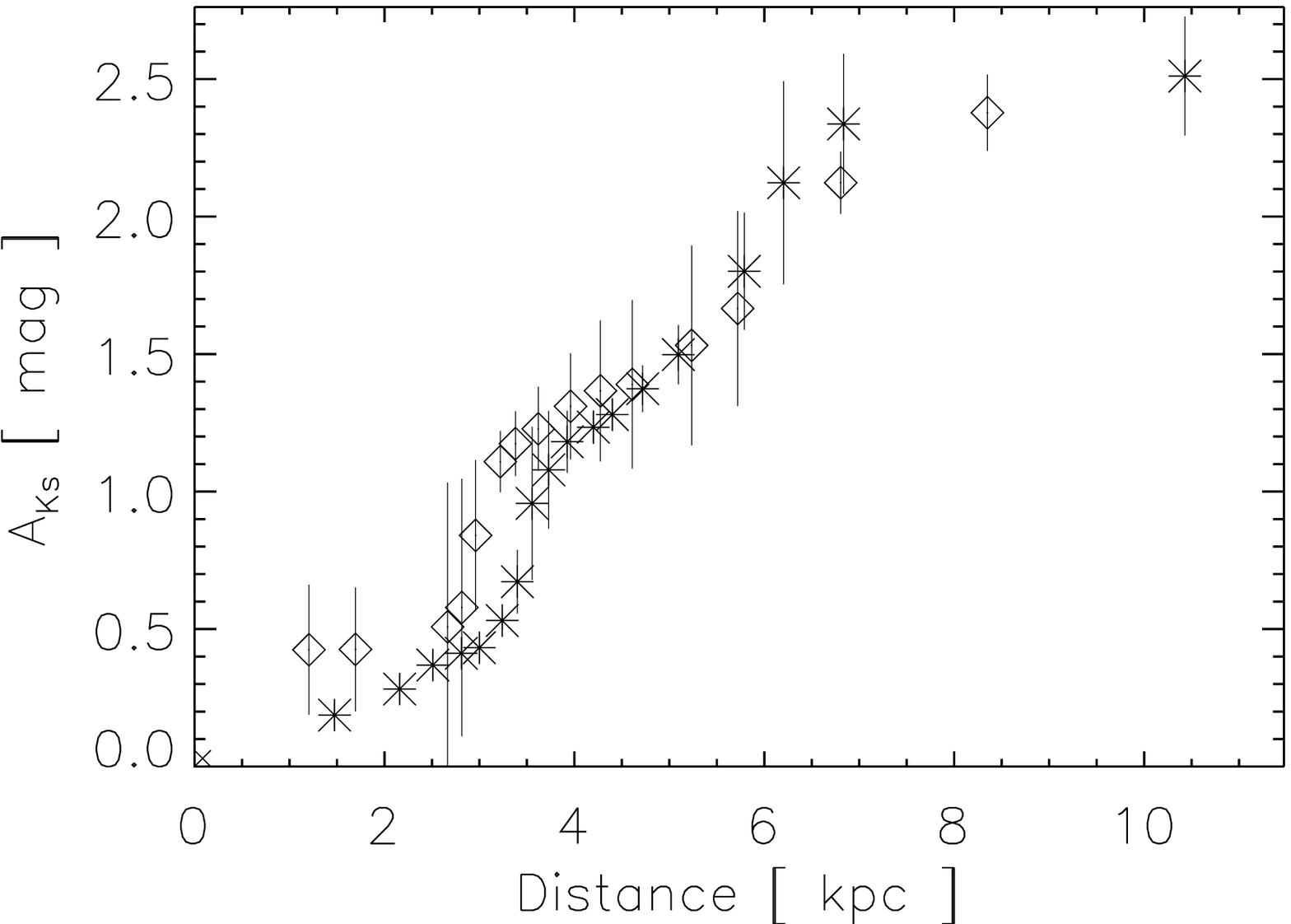,width=0.8\linewidth}} 
      {\epsfig{file=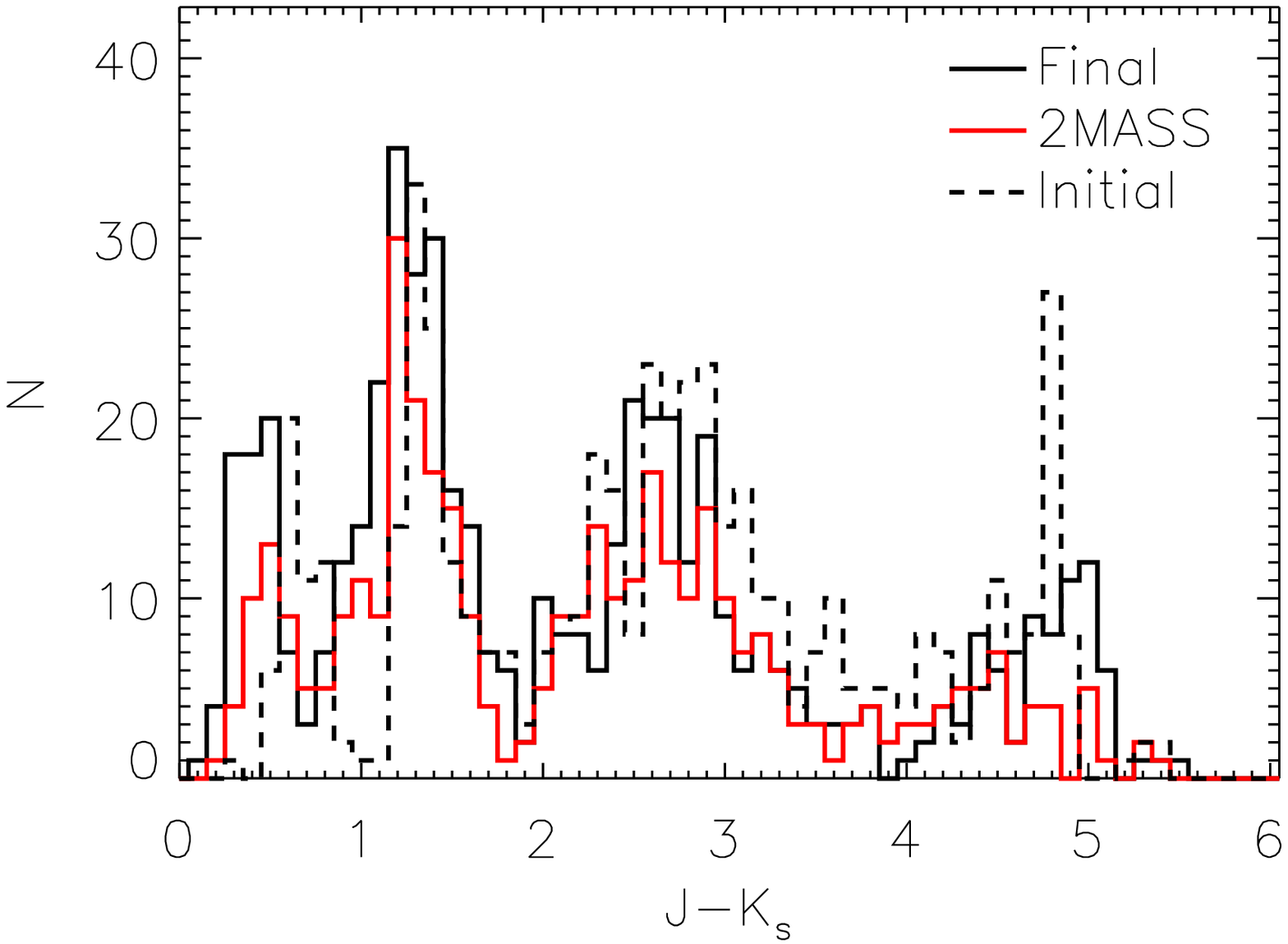,width=0.8\linewidth}} 
      \caption{Initial and final 
	results of the extinction calculation for the field $(l,b)=(345,0)$. 
      {\bf Top:} $A_{Ks}$ as a function of distance. Each diamond represents a bin from 
the initial calculation, the asterisks 
represent the bins in the final result and the error bars represent the mean absolute deviation of the extinction in the bin.
      {\bf Bottom:} $J{-}K_{s}$ colour distribution of 2MASS and adjusted Galactic model. 
      The red solid line represents the 2MASS data, the bold solid line represents 
      the final result and the broken line shows our initial calculation}
      \label{fig:example}
\end{center}
\end{figure}
{\changes This method supposes
that the observed and simulated stars in a given bin are at similar distances. As we start with a simple
distribution for the extinction this may not be the case, initially. 
However, each successive iteration improves the extinction estimation and decreases the 
difference between the observed and modelled colour distributions. 
Assuming that the spatial density of the giants is well modelled in the Galactic model, and that the extinction
is uniform across the field (\S\ref{sec:cloud_sub}),
the assumption that the modelled and observed colour bins represent stars at similar distances
is justifiable.
}

{\changes
As an example, the initial (after one iteration) and final results for a sample field are displayed in Fig.\ref{fig:example}. 
The initial extinction (diamonds) is  significantly different from the final extinction (asterisks),
with nearly a 1 kpc difference in the location of the sharp rise in extinction at $\sim$3 kpc. 
This difference results in a poorer fit to the observed colour distribution and therefore
a large value for the $X^2$ statistic (Eq.\ref{eqn:X2}). 

We may summarise the various steps of the method as follows:
\begin{enumerate}
\item
Assume simple distribution for the extinction \S\ref{sec:model_ext}, or
apply extinction from last iteration 
\item
Remove dwarf stars and cut observations and model at magnitude limits
\item
Sort both catalogues by increasing $J{-}K_{s}$ colour
\item
Group simulated stars into colour bins such that the distance of each  bin $\ge$ distance of previous bin
\item
Group observations into same number of bins as simulated star bins, each bin having the same relative number of stars
\item
Compute extinction correction using Eq.\ref{eqn:dAk}, bin by bin
\item
Compute $J{-}K_s$ histogram and $\chi^2$ statistic
\item
Continue until the $\chi^2$ statistic reaches a minimum
\end{enumerate}
}


\section{Results}

The full results are listed in Table 1, only available in electronic form
at the CDS via anonymous ftp to cdsarc.u-strasbg.fr (130.79.128.5)
or via http://cdsweb.u-strasbg.fr/cgi-bin/qcat?J/A+A/. Each row of Table 1 contains the 
information for one line of sight: Galactic coordinates along with  
the measured quantities for each bin ($A_{Ks}$, distance and respective uncertainties).
All rows also contain
the corresponding $\chi^2$ statistic to indicate the quality of the fit between the 
modelled and observed $J-K_s$
histograms.

\label{sec:results}
\begin{figure*}
  \begin{center}
  \leavevmode
      {\epsfig{file=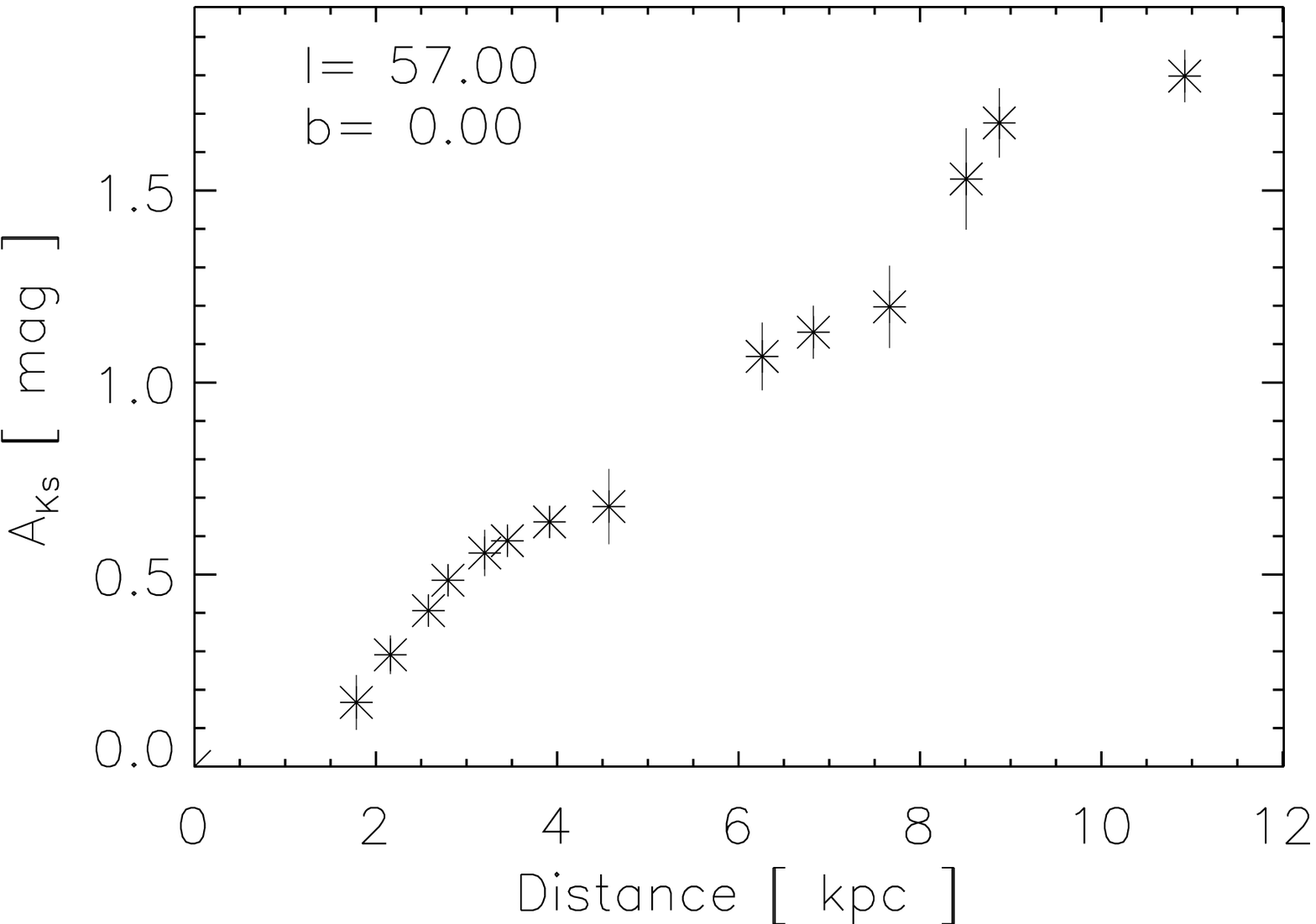,width=0.33\linewidth}} 
      {\epsfig{file=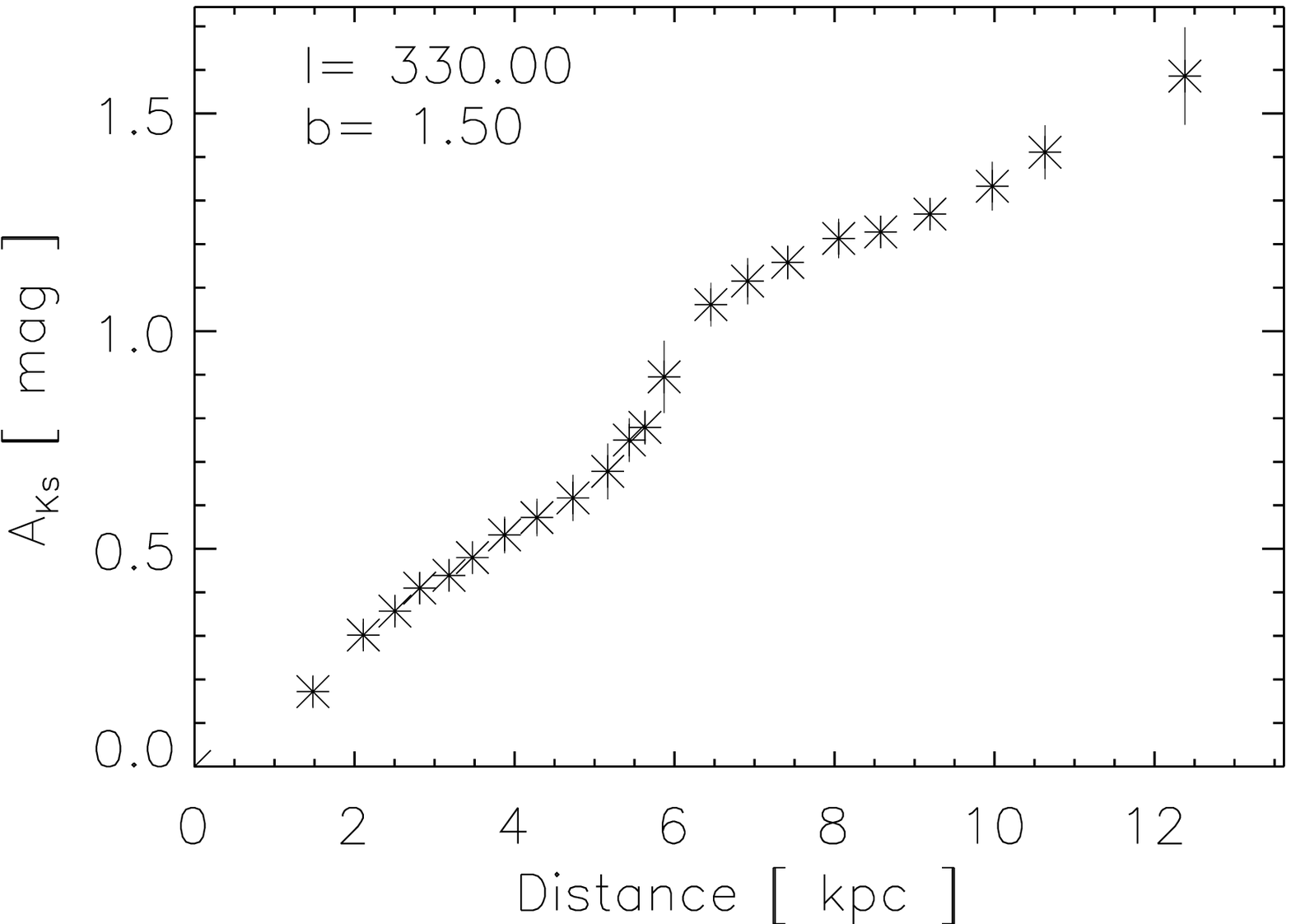,width=0.33\linewidth}} 
      {\epsfig{file=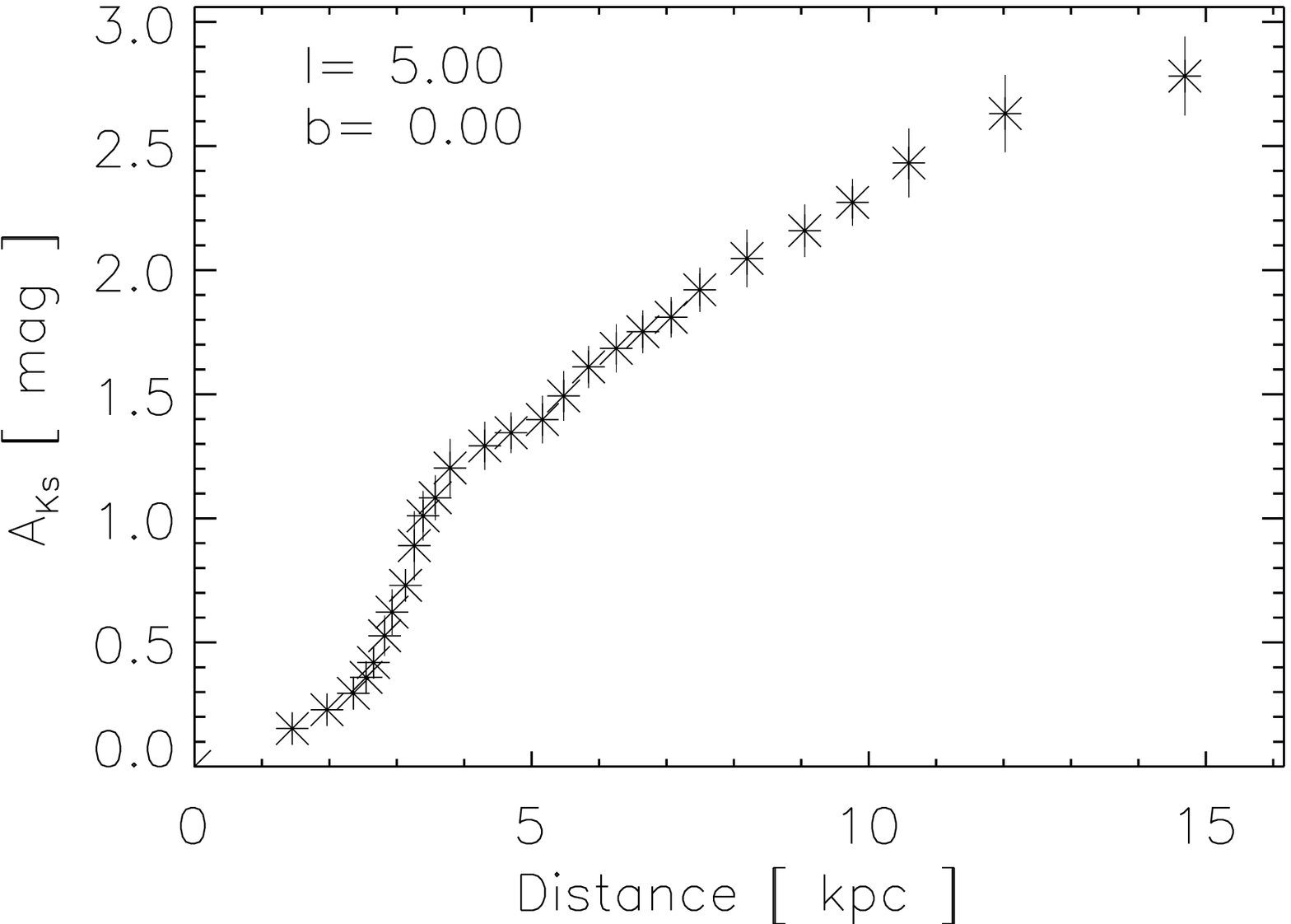,width=0.33\linewidth}}       
      {\epsfig{file=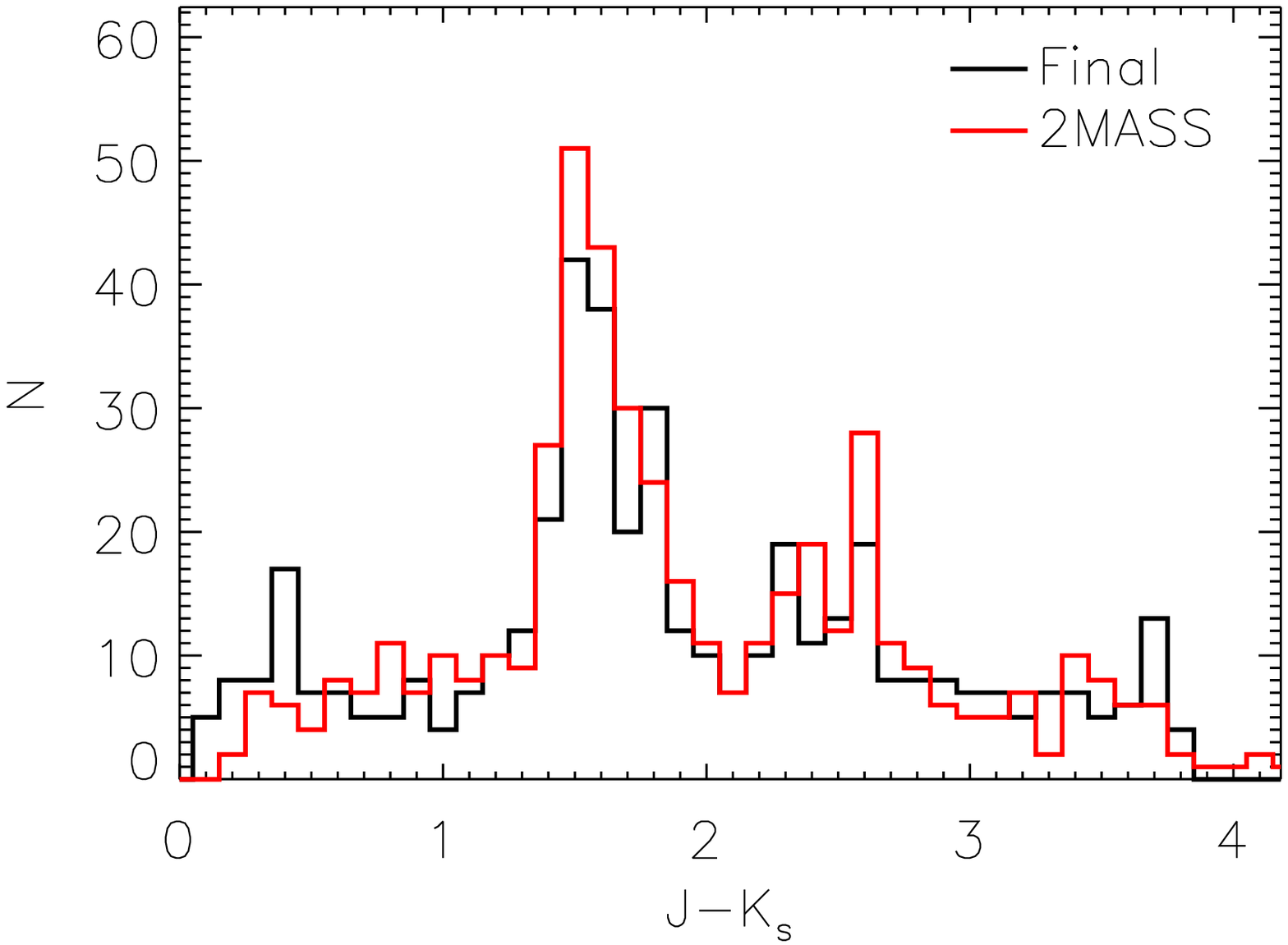,width=0.33\linewidth}} 
      {\epsfig{file=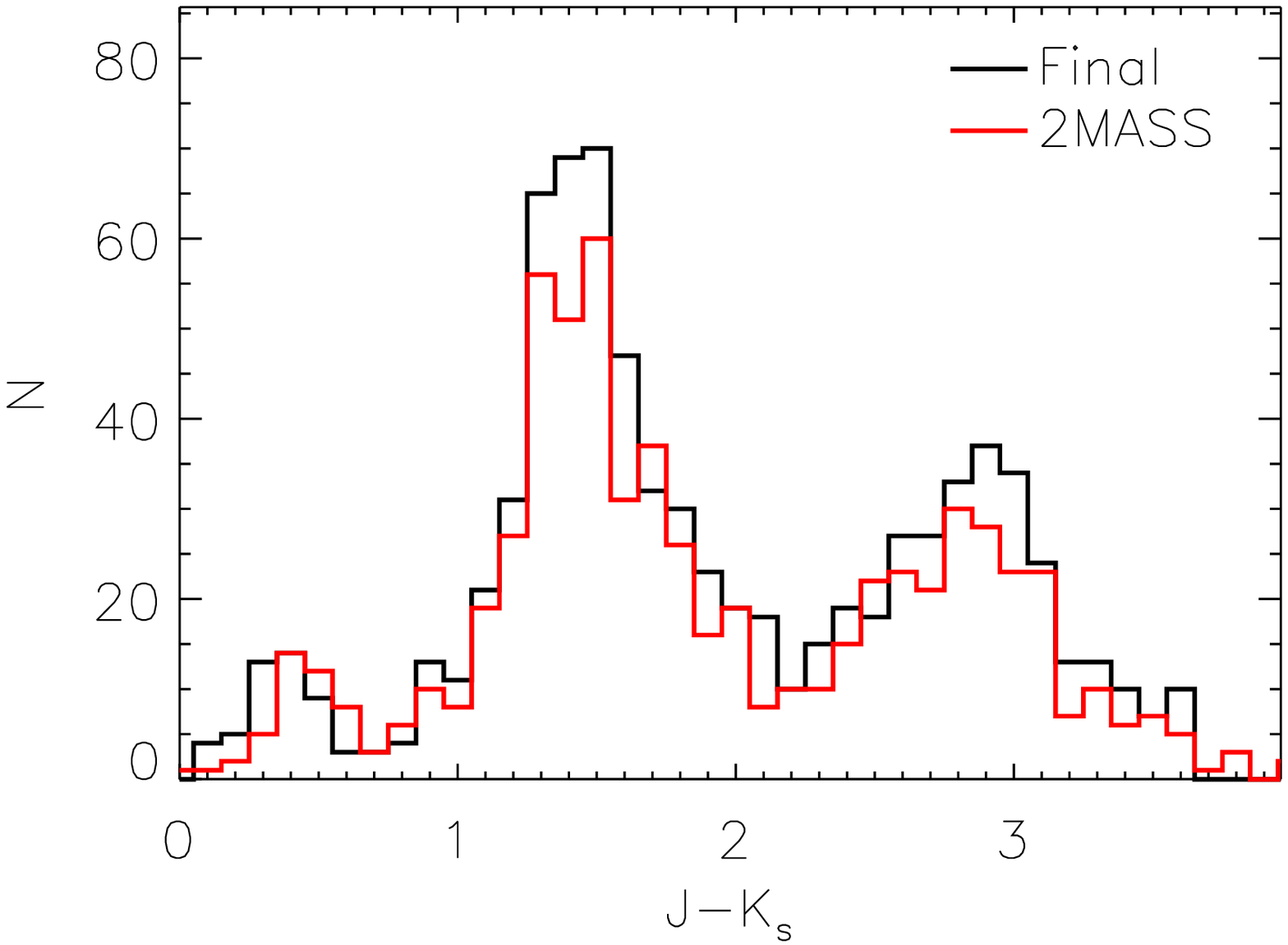,width=0.33\linewidth}} 
      {\epsfig{file=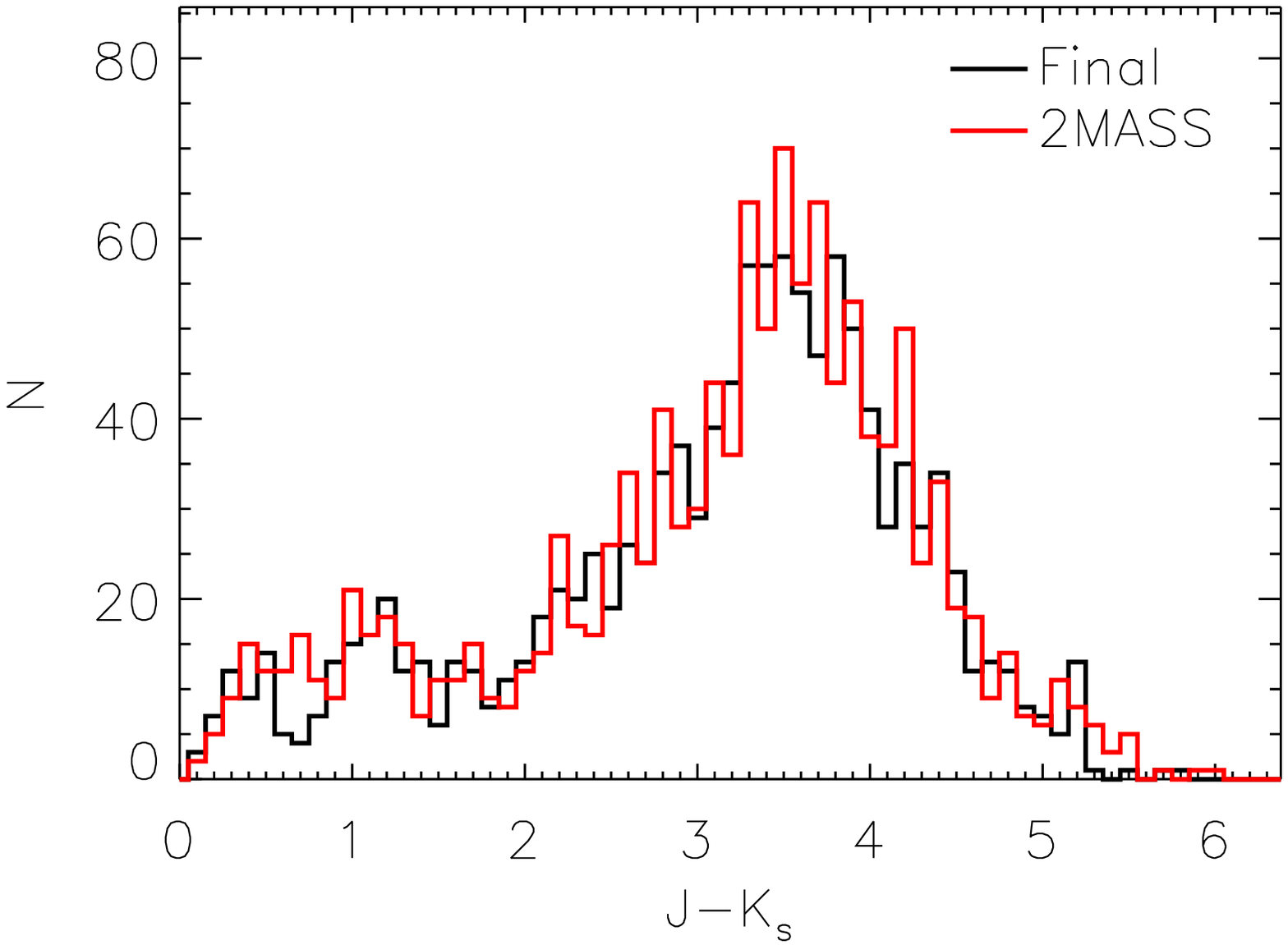,width=0.33\linewidth}} 
      \caption{Results for the three fields at galactic coordinates $(l,b) = (57,0), (330,1.5)$ and $(5,0)$. 
	{\bf Top}: $A_{Ks}$ vs. distance. 
	Each asterisk represents a bin from our method, and the error bars are the mean absolute deviation from the median
	extinction in the bin.
	{\bf Bottom}: Corresponding $J{-}K_s$ histogram of the model and the observations. The red line represents
	the 2MASS observations and the black line is the result of our method } 
  \label{fig:ex_results}
\end{center}
\end{figure*}
\subsection{Results for individual lines of sight}
The results for three example lines of sight are presented in Fig. \ref{fig:ex_results}: $(l,b)=(57,0),(330,1.5)$ and $(5,0)$.
The top row presents the distribution of extinction along the line of sight;
 each asterisk represents one bin from our results and the error bar is the absolute mean deviation of the 
$A_{Ks}$ from the
median of all the {\changes modelled} stars in the bin. 
{\changes Each simulated star at a given distance is attributed an extinction by 
 performing a linear interpolation between the two neighbouring bins}.
In order to test the extinction law thus determined, 
we compare the $J{-}K_{s}$ histogram from the 2MASS observations to the $J{-}K_{s}$ histogram resulting
 from the application of our extinction to the Galactic model.
The histogram for each line of sight is presented in the bottom row of  Fig. \ref{fig:ex_results}.
As can be seen from these three examples, our method works very well in areas of high extinction. 

\subsection{Two dimensional extinction maps}

By combining many lines of sight we are able to construct a three dimensional array of extinction values
in the Galaxy. However, the values of $A_{Ks}$ along each line of sight are not distributed uniformly.
If we want to know the extinction along a plane at a certain heliocentric distance we must first 
interpolate in between the values in order to get a regular grid. In what follows we have used
 linear interpolation between successive bins. 

\subsubsection{Comparison with two dimensional extinction maps}

Two dimensional extinction maps towards the Galactic centre were computed by 
 \citet{schultheis1999} ($|l|\le8\degr$, $|b|\le1.5\degr$) and \citet{dutra2003} ($|l|\le5\degr$, $|b|\le5\degr$), 
hereafter S99 and D03 respectively.
The method of S99 is based on the comparison of AGB isochrones
to theoretical unreddened isochrones \citep{bertelli1994}, using DENIS data, whereas D03 compute the extinction 
by fitting the upper giant branch of $(K_{s}, J{-}K_{s} )$ 
colour magnitude diagrams to a de-reddened upper giant branch mean locus
of previously studied bulge fields.

Although they used near-infrared data as we have, our approach provides a different method
for extracting the extinction information from the data.

In order to compare our map with theirs, we first smooth their maps to our resolution of 15\arcmin $\times$ 15\arcmin.
The S99 method yields results in 
$A_{Ks}$ but they were published in $A_{V}$; we use the extinction law from \citet{glass1999}, 
$A_{Ks}=0.089A_{V}$,
 as they did to reconvert the $A_{V}$ to $A_{Ks}$. The D03 map was published in $A_{K}$
so we convert it to $A_{Ks}$ using the conversion factor used in their article $(A_{K}=0.95A_{Ks})$.
Finally, we integrate our extinction to the centre of the Galaxy, assumed to be at 8.5 kpc from the Sun.

\begin{figure}
  \begin{center}
  \leavevmode
      {\epsfig{file=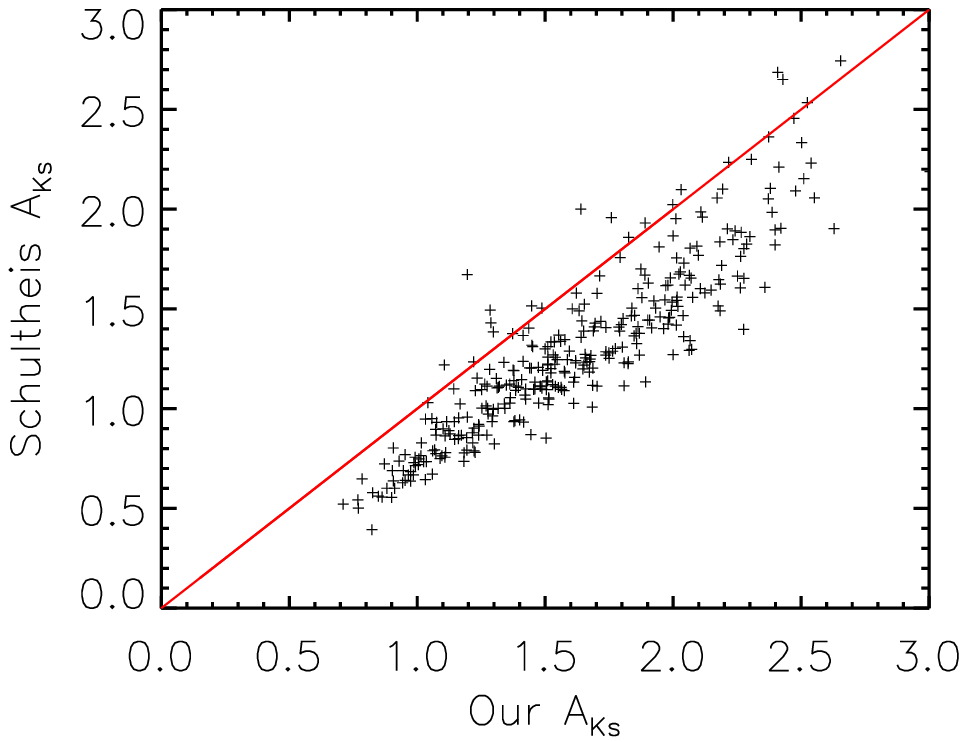,width=0.48\linewidth}} 
      {\epsfig{file=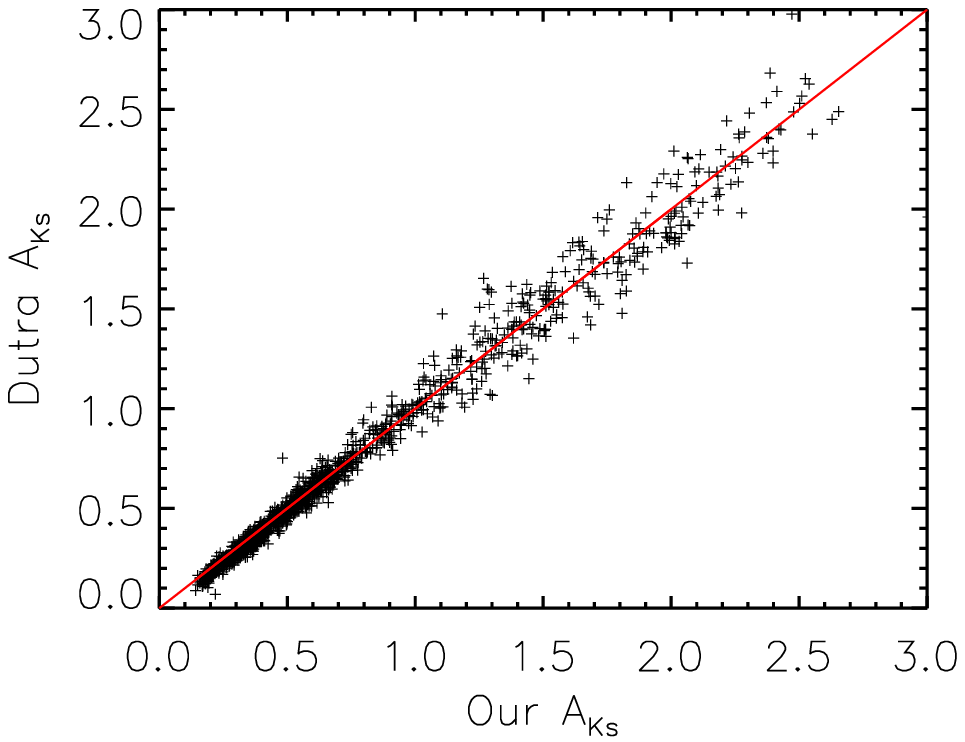,width=0.48\linewidth}} 
  \caption{Comparison between each pixel of
    our extinction map  and that of \cite{schultheis1999} (S99), for $|l|<8\degr$, $|b|<1\degr$,  
    and \cite{dutra2003} (D03), for $|l|\le5.0\degr$, $|b|\le5.0\degr$.
    {Left}: $A_{Ks}$ of our map vs. that of S99.  
    {Right}: $A_{Ks}$ of our map vs. that of D03. 
    For both figures, the straight line represents the identity function. }

  \label{fig:histo_mathias}
   \end{center}
\end{figure}

The results of this comparison are displayed in  Fig. \ref{fig:histo_mathias}, which
shows a comparison between our results and the S99 map (left) and D03 map (right) at each pixel. 

Although the large scale structure of all three maps are very similar there is a systematic difference of $\sim0.2-0.3
$ $A_{Ks}$ between our map and that of S99. 
{\changes 
This is not surprising as they use the extinction law from \cite{glass1999}, while we use the results of \cite{mathis1990},
 resulting in a different ratio for the $K_s$ band to $J$ band extinction ($A_{Ks}/A_J$). This ratio is used
 in the derivation of  Eq.\ref{eqn:dAk} and therefore has a direct impact on the extinction determination. 

The extinction law in the infrared can differ significantly
 from one author to the next; the value we have adopted has been chosen so as to be
 consistent with the extinction law used for the Galactic model \citep{mathis1990}. 

By combining  
the \cite{mathis1990} law,  $A_{K}/A_J= 0.382$ with the relation $A_K=0.95A_{Ks}$ \citep{dutra2002} 
we obtain $A_{Ks}/A_J= 0.402$, very close to the 
recent value ($A_{Ks}/A_J=0.400$) from \cite{indebetouw2005}. 
However, this
value is significantly different from \cite{glass1999} ($A_{Ks}/A_J=0.347$), which is 
 closer to the value recently measured by \cite{nishiyama2006}, $A_{Ks}/A_J=0.331$, 
in their study of the Galactic bulge. 

The application of the \cite{mathis1990} law to the S99 results effectively removes the systematic shift
 seen in Fig.\ref{fig:histo_mathias}.
D03 already use the same value for  $A_{Ks}/A_J$ as we do;  
the correlation between our map and theirs is very good.
}
The slight departure from the identity function at low
extinction, comes from the poor sensitivity of the $J{-}K_{s}$ colour index at these relatively low extinctions 
(\S\ref{sec:maxmin}). 
{\changes These low extinctions do not exist in the S99 map as the area covered by their map does not rise 
much above $|b|\sim1.5$; in addition to the more central regions of the Galaxy, 
the D03 map also probes higher latitude, lower extinction lines of sight.
}

\subsection{Comparison with velocity integrated CO emission}
 
 \begin{figure*}
  \begin{center}
    \leavevmode
    \epsfig{file=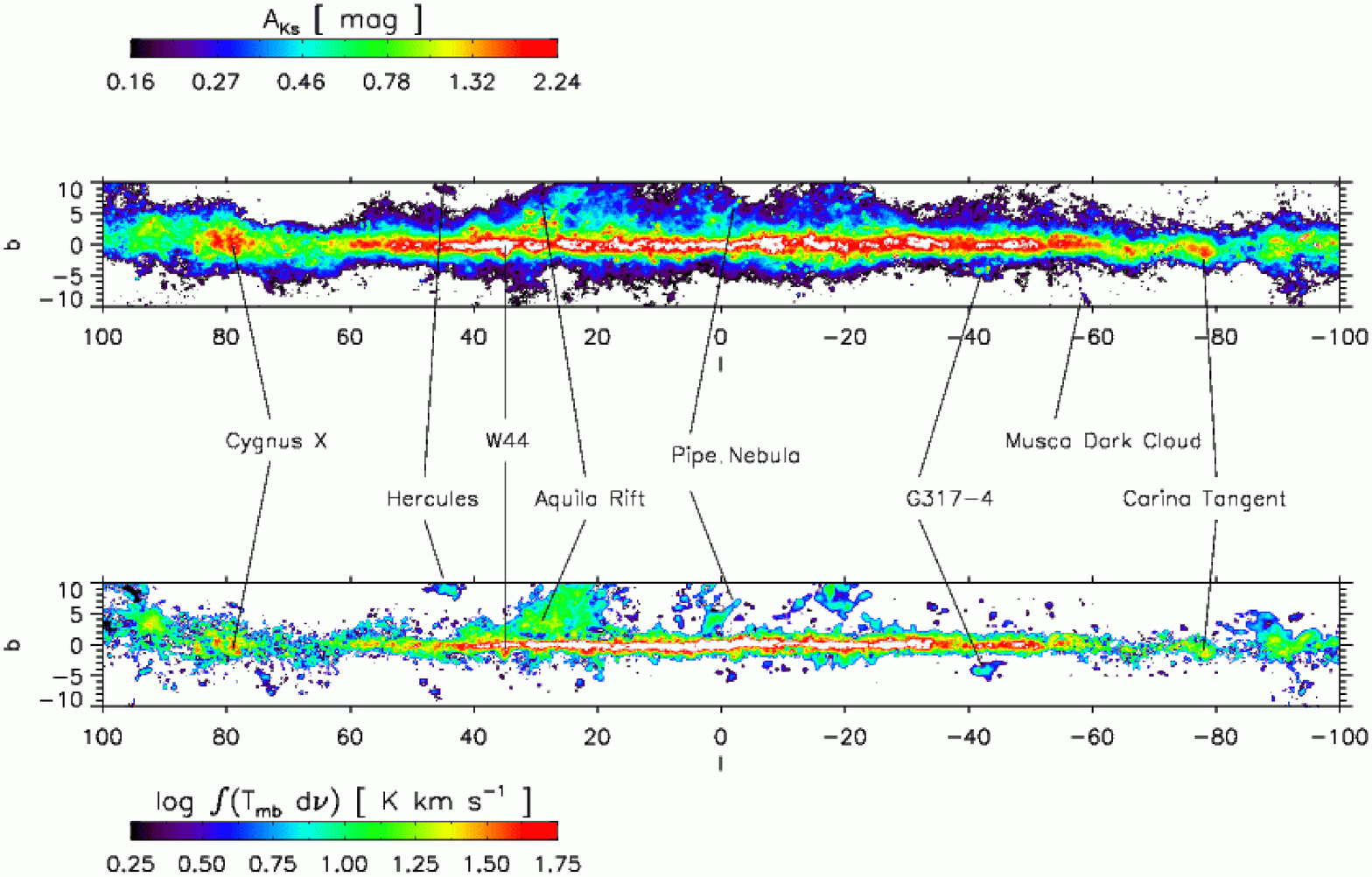,width=1.3\linewidth,angle=270}
   \end{center}
  \caption{{\bf Top}: Total extinction integrated along the line of sight. 
    {\bf Bottom}: CO velocity integrated spatial map by \cite{dame2001}. The units of the two maps are different;
they are put here to compare their respective interpretation of the large scale structure of the Galaxy.
  The coordinates are expressed in degrees $(l,b)$.}
  \label{fig:centre_map}
\end{figure*}

\cite{dame2001} have created a large scale CO survey of the entire Milky Way, using the rotational transition
1-0 of the CO molecule at 115 GHz, from new observations and existing CO surveys. 
By comparing 
the low velocity (local) CO emission with an optical panorama of the inner Galaxy,
they conclude that there are few dark interstellar clouds where CO is not present or
where the temperature is so cold that the CO molecule fails to 
radiate at 115 GHz. 
{\changes Galactic CO emission traces only dense regions of the ISM whereas our map also traces
the diffuse component. Nevertheless
we expect that both maps should trace the 
large scale distribution of interstellar matter in our Galaxy. }

Fig. \ref{fig:centre_map} shows the total extinction detected for each line of sight using our method as well
as the distribution of CO as presented in \cite{dame2001}, where the two maps overlap. 
The resolution of the extinction map is 15\arcmin x 15\arcmin; the resolution of the CO map varies from
 $9\arcmin$ to $30\arcmin$, the
region that we present here is mostly at the higher resolution of 9\arcmin. The main structures that appear
in both maps are indicated. 
The Musca Dark cloud, $(l,b)=(-59,-9)$, that appears in our map is not in the \cite{dame2001} map
as this region was not observed in their composite survey.

\subsection{Three dimensional extinction}

{\changes 
In order to visualize the distribution of extinction elements in three dimensions, we divide 
the extinction between subsequent bins by the distance between them. The resulting map, in units of $\delta A_K$ kpc$^{-1}$,
gives us  a better idea of the location  of obscuring dust along the line of sight. We assume a constant 
$\delta A_K$ kpc$^{-1}$ in between bins, as we have no information on the
distribution of dust between bins. This results in an overestimation of the line 
 of sight length of clouds for lines of sight with large distances between bins.}

\begin{figure*}
  \begin{center}
     \leavevmode
     \epsfig{file=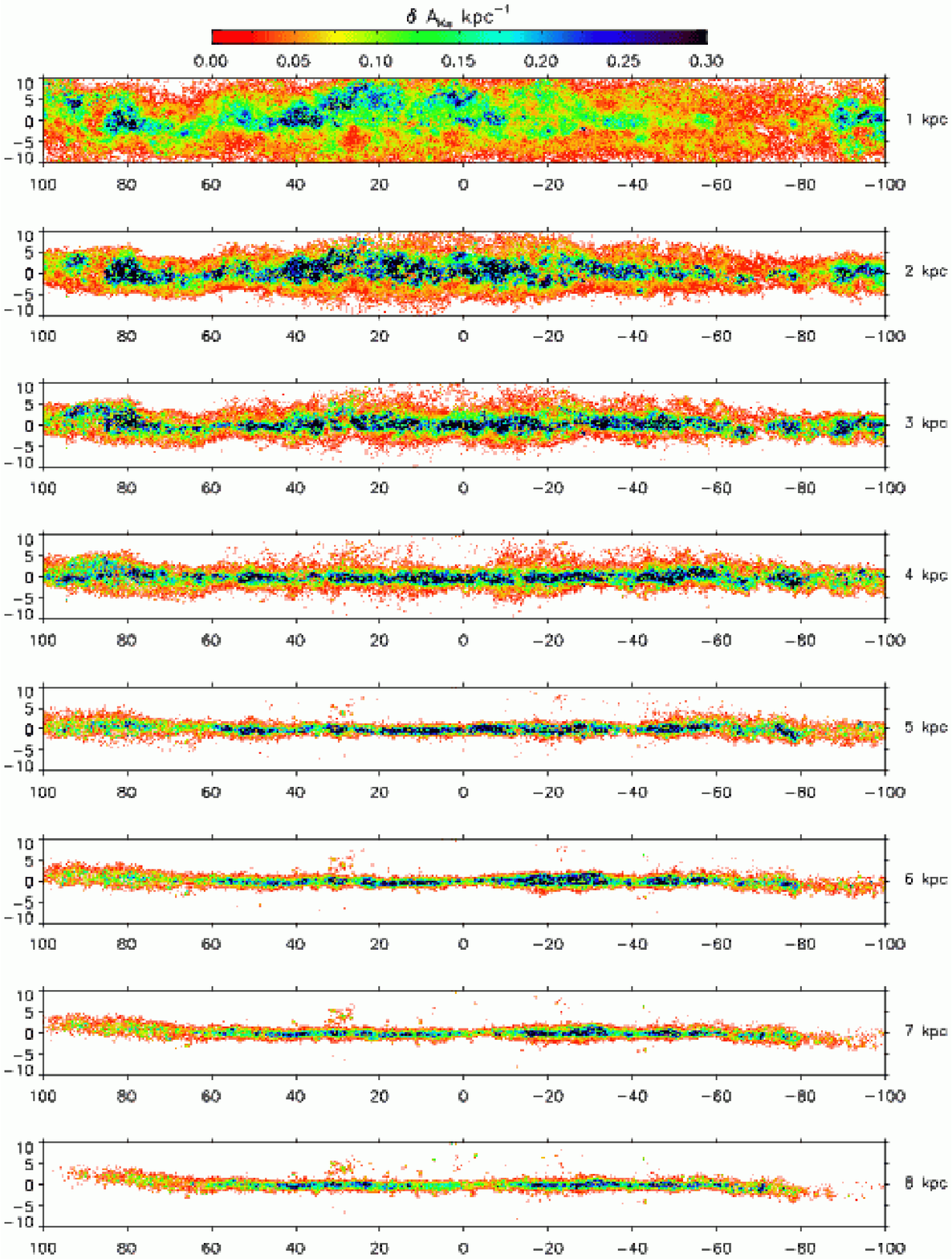,width=\linewidth}
   \end{center}
  \caption{The local extinction at 1 kpc intervals from the sun.
The uppermost image
is at 1 kpc, the bottom one is at 8 kpc.
The x axis is in Galactic longitude, the y axis in Galactic latitude. 
The solid line indicates the mean position of the plane, as given by
the stellar warp formula in the Galactic model.
Different structures
are identified in the text (\S\ref{sec:3D_ext}).} 
  \label{fig:slices}
\end{figure*}

These maps are presented in two forms: 
maps of the full area covered so far by our method at various distances  (Fig. \ref{fig:slices},
 and fully described in \S\ref{sec:3D_ext}), as well as
a view of the 
Galactic plane from above (Fig. \ref{fig:birdseye}, referred to in \S\ref{sec:birdseye}).

\subsubsection{Distance slices}
\label{sec:3D_ext}

The three dimensional extinction towards the inner Galaxy 
is presented in Fig. \ref{fig:slices}.
{\changes Each image
shows the average local extinction at distance intervals of 1 kpc. 
The scale is the same for each image and runs from  0 to 0.3 $A_{Ks}$ kpc$^{-1}$}. 
The solid line superimposed at each distance shows the stellar warp as implemented in the Galactic model.

One obvious feature in 
the images
is the geometrical effect of distance; high latitude extinction is mostly local so 
the disc is seen to shrink as we go further away from
the Sun. In the first couple of slices, the local features mentioned earlier are visible: the Aquila Rift (30{\degr},
{4\degr}) and the
Local Arm towards Cygnus X (80{\degr},0{\degr}).

The giant molecular cloud (GMC), 
and supernova remnant, W44 is visible in the second and third slices (2kpc \& 3kpc) at (35\degr,0\degr).
{\changes 
At distances between 3 and 5 kpc the tangent to the Carina arm can be seen at $\sim -78\degr$.
}

After 6 kpc, a lack of absorbing material is noticeable towards the Galactic bulge, an effect noted by \cite{glass1987}
who found that the bulge has little dust compared to its CO emission. 
At these distances another feature of the disc which starts to appear is the Galactic warp. It becomes
more pronounced at large absolute longitudes as we move farther out from the Sun. 

{\changes
Finally, the lack of line of sight resolution for certain fields is obvious at $(l,b) \sim (30,5)$ and at
distances greater than 3 kpc. Due to the fact that we use 
linear interpolation between successive bins and therefore do not
assume any structure for the distribution of the dust in between bins, 
the large distance between successive bins
 creates an unrealistically long structure along certain lines of sight. 
The extinction estimation itself is 
still assumed to be correct, within error.
}

\subsubsection{The Galactic plane}
\label{sec:birdseye}

\begin{figure*}
  \begin{center}
    \leavevmode
	{\epsfig{file=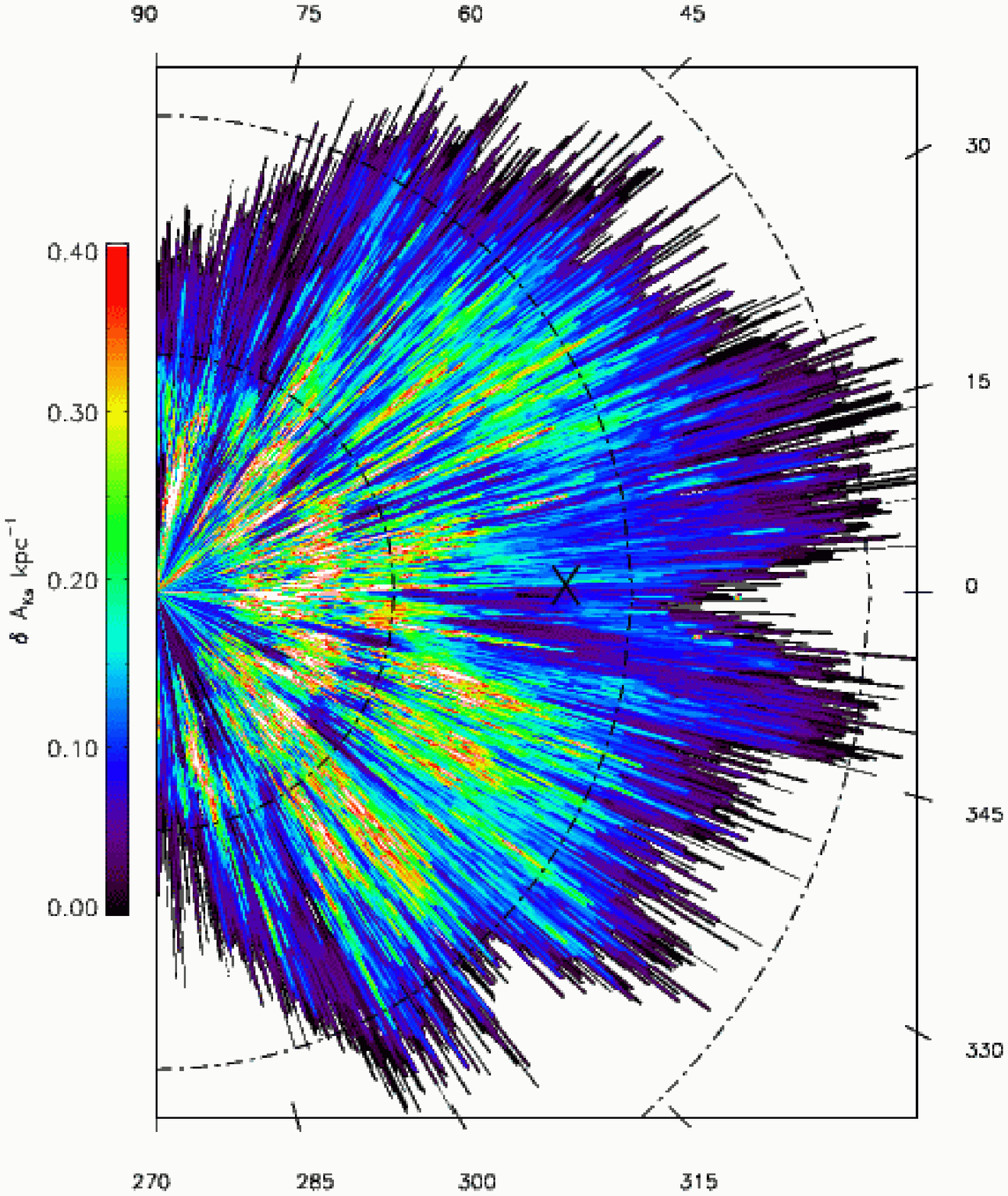,width=\linewidth}}
  \end{center}
  \caption{Location of absorbing dust in the Galactic Plane ($|b| \le 0.25$). 
The Galactic centre, assumed to be at 8.5 kpc from the Sun, is marked with an 'X'; 
the Sun is located in the middle on the left side of the plot.
The dotted lines are placed every 5 kpc from the Sun. Ticks on the border indicate Galactic longitude.}
  \label{fig:birdseye}
\end{figure*}

By creating a map showing the local extinction in $\delta A_{Ks}$ kpc$^{-1}$,
we are able to visualise the location
of absorbing matter in the inner Galaxy. 
A view ``from above'' is displayed in Fig. \ref{fig:birdseye}
which is the average $\delta A_{Ks}$ for $|b| \le 0.25$ towards the inner Galaxy. 
The Sun is located
at the middle left of the graph, and the Galactic centre is marked by an 'X'.

Many structures are visible in this figure, such as the local arm
extending outwards from the Sun at $l\sim$80{\degr}. 
{\changes Two tangents to the Sagittarius-Carina arm
are also visible, extending from the Sun in two directions at $l\sim 280\degr$ and 
$l\sim50{\degr}$}.
The outer Perseus arm is above the top edge of the map and is therefore  not readily identifiable. 
{\changes The inter-arm region between the Centaurus-Crux and Norma arms is visible at $\sim$ 4.0 kpc from the Galactic
Centre.
After the Norma arm,}
centred on the Galactic centre is a large void of dust apart from an elongated structure
which may well indicate the presence of dust in the Galactic bar, which we discuss below.

\subsection{Galactic parameters}
We concentrate on determining large scale characteristics of the 3D dust 
distribution from our 3D extinction map. The mean scale height, as well as 
the shape of the warp are measured. An estimate of the central dust bar 
orientation and size is also given.

\subsubsection{ISM disc scale height}

{\changes
\begin{figure}
  \begin{center}
    \leavevmode
    \epsfig{file=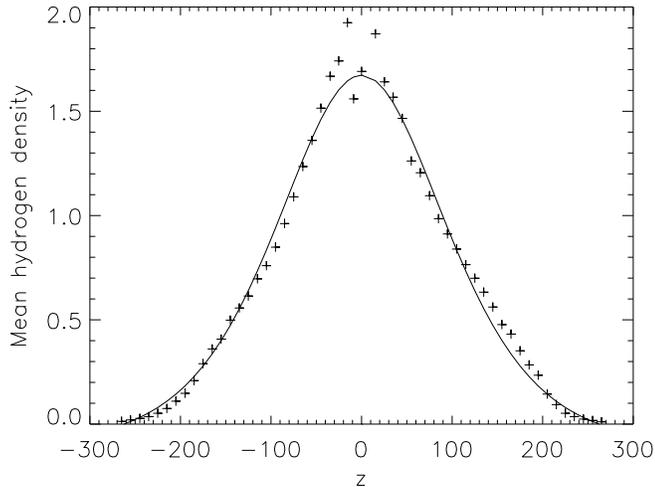,width=\linewidth}
    \caption{Variation of the density of absorbing matter as a function of distance from the Galactic plane.}
  \label{fig:sc_height}
  \end{center}
\end{figure}

Our 3D extinction map allows us to calculate the scale height
of the dust component of the ISM
The
extinction between two successive bins 
was converted to a hydrogen column density using the results from \cite{bohlin1978}; this
column density was then divided by the distance between the bins to obtain a hydrogen density per volume. This supposes
that the metallicity is the same as in the solar neighbourhood;
\cite{chiappini2001} show that the abundance gradient in the Milky Way is relatively flat between 4 and 10 kpc
from the Galactic centre but that it
steepens sharply in the outer disc. To minimise the effect of any such gradient on our calculation,
 we avoid all areas within 4 kpc of the Galactic
centre and beyond the solar circle. 
We then calculated the mean hydrogen density at every 10 pc from the Galactic plane.

In order to measure the scale length of the detected interstellar matter, we tested two profiles :
\begin{equation}
  \rho=\rho_0{\rm sech}^2(z/h_z) 
\end{equation}
\begin{equation}
  \rho=\rho_0{\rm e}^{(-|z|/h_z)} 
\end{equation}
where $h_z$ is the scale height of the ISM and $\rho_0$ the hydrogen density in the Galactic plane.
The use of the sech$^2$ profile results in a better fit and is shown in Fig.\ref{fig:sc_height}.
The exponential
 gives $h_z=134^{+44}_{-11}$ pc 
and $\rho_0 = 2.5 \pm 1.3$ atoms cm$^{-3}$, whereas the sech$^2$ profile yields  $h_z=$ \scheight and $\rho_0 = 1.8 \pm 1.0$ atoms cm$^{-3}$. 
Both \cite{malhotra1999} and
 \cite{nakanishi2003} find that the scale height of the HI disc in the Milky Way varies as a function of distance; for both 
the mean value between 4 and 8 kpc is of the order of $\sim160$ pc.
The dust model
of \cite{drimmel2001}, adjusted to the FIR emission, includes a  scale height of $134.4\pm8.5$ pc.
 We were not able to detect any significant flaring of the interstellar matter disc within
the Galactocentric radii we tested.
}

\subsubsection{Galactic warp}
\label{sec:warp}

  \begin{figure}
  \begin{center}
    \leavevmode
    \epsfig{file=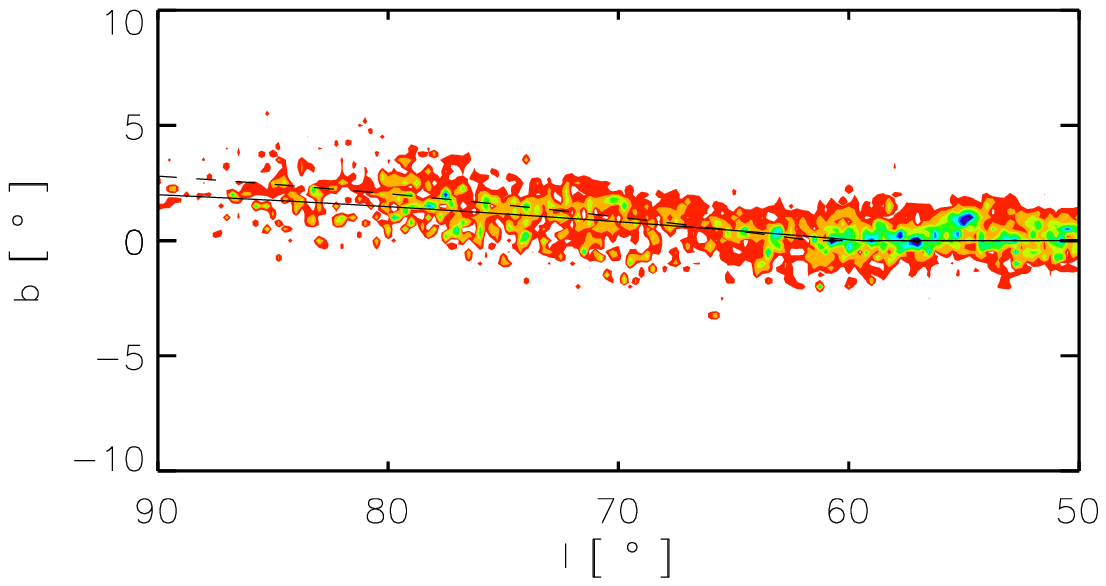,width=\linewidth}
    \epsfig{file=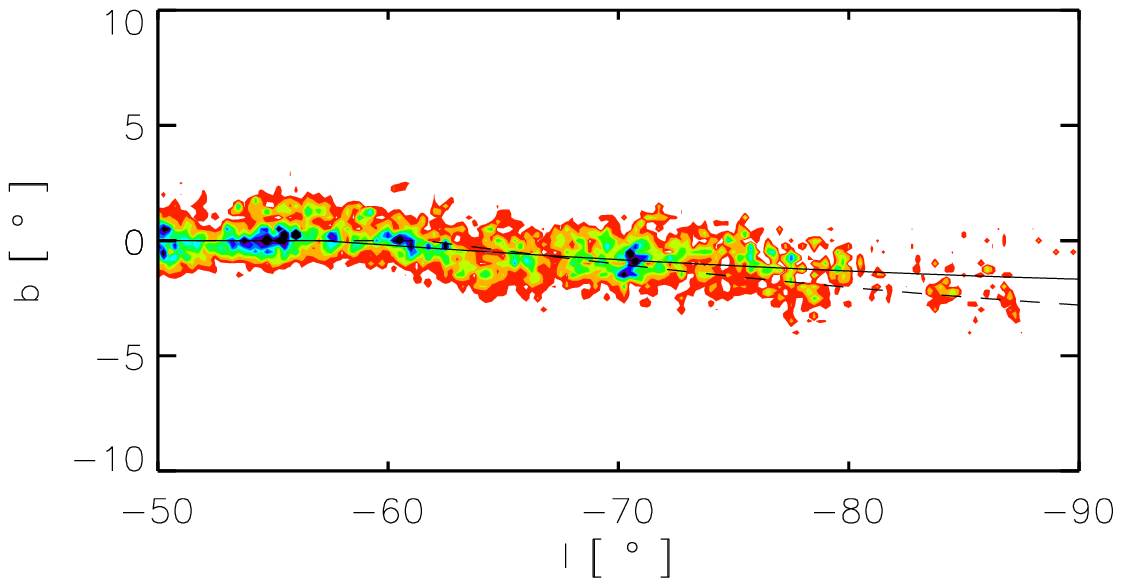,width=\linewidth}
    \caption{Local extinction at 8 kpc from the Sun for positive longitudes (top) and negative longitudes (bottom). 
      The stellar warp as modelled in the Galactic model is shown
      as a dashed line; the solid line shows the best fit to our extinction data (\S\ref{sec:warp}). 
      The colour scale is the same as in Fig.\ref{fig:slices}}
  \label{fig:warp_test}
  \end{center}
\end{figure}
{\changes
The Galactic warp is obvious in figure \ref{fig:slices}, especially at large absolute longitudes and heliocentric distance. 
The magnitude of the  warp in the Galactic model is calculated using Eq. \ref{eqn:warp} :
\begin{equation}
\label{eqn:warp}
  z_{\rm warp} = \gamma * (R - r_{\rm warp})*\cos(\theta-\theta_{\rm warp})
\end{equation}
where $z_{\rm warp}$ is the perpendicular shift from the mean $b=0\degr$ plane, 
$R$, $\theta$ and $z$ are Galactocentric coordinates, $\gamma$ is the slope of the displacement of 
the mid plane with respect to the plane of symmetry, 
 $r_{\rm warp}$ is the radius after which
the warp starts and $\theta_{\rm warp}$ is the angle of maximum warp. Using our results, we are able to 
locate the $z$ with the highest density for each $(r,\theta)$ by calculating the best fit with Eq.\ref{eqn:warp}.
We find that the warp is not symmetrical; for positive longitudes we find $\gamma=\slopepos$, $\theta_{\rm warp}=\angpos$ and 
$r_{\rm warp}=\radpos$ and for negative longitudes we find $\gamma=\slopeneg$, $\theta_{\rm warp}=\angneg$ and 
$r_{\rm warp}=\radneg$. 
Fig.\ref{fig:warp_test} shows the local extinction per kpc at a heliocentric distance of 8 kpc, overlaid 
with the warp parameters of the Galactic model
(dashed line) and with the above parameters (solid line). These values correspond to an absolute vertical shift
of the plane at $R=12$ kpc of \warpmaxneg at both negative and positive longitudes, significantly
 smaller than the modelled stellar warp
which predicts a vertical shift of $\sim$ 650 pc at $R=12$ kpc.
 These new values for the dust warp are slightly higher
than those of \cite{drimmel2001} ($\gamma =0.0728$, $r_{\rm warp}=6.993$ kpc), who predict a maximum 
vertical shift of 364 pc at 
$R=12$ kpc. This shift is different for the stellar warp, which they find to be higher.
\cite{lopez2002} use a different formulation to model the warp and find the maximum vertical shift of the stellar 
warp at $R=12$ kpc to be 547 pc, in good agreement with the gas warp found by \cite{burton1988}.

\cite{nakanishi2003} find that
the Galactic HI disc is warped and asymmetric, with a larger shift at positive longitudes,
 as we do.
They report that the warp is at a maximum for $\theta=80\degr$ and $\theta=260\degr$, and that
 the warp starts at $R=12$ kpc, contrary to our values at around $R=8$ kpc. 
At positive longitudes the warp continues to reach a maximum displacement from the
 $b=0\degr$ plane of 1.5 kpc at $R=16$ kpc, resulting in a slope of 0.375. 
At negative longitudes they find that the warp is less severe, 
resulting in a shift of 1.0 kpc below the $b=0\degr$ plane at $R=16$ kpc, a slope of 0.25. Their values
 for the gradient of the warp are significantly higher than ours, however our map does not have information out to these 
large galactocentric distances and so we are not able to constrain the large scale attributes of the interstellar warp in the
 outer Galaxy.

Most recently \cite{levine2006} have shown that the outer HI warp can be
described by a superposition of the three low mode (m = 0, 1 and 2)
vertical harmonics of the disc, one of which (m = 1 corresponding to
the integral-sign shape) dominates the warp for $R\la15$ kpc.
It seems that
it is this mode that we see in the dust from our study, which does not
extend far enough into the external regions of the Galaxy to detect the
other modes. They find that this mode is approximately linear for $10 \la R \la 15$ kpc and that
it has a slope of 0.197, significantly higher than our value.

From these comparisons one sees that the warp is present in all
galactic components (dust, gas and stars),all with the same node
position and all asymmetric. However, the amplitude of the warp seems to
 depend slightly on the component one looks at: the dust warp \citep[our study;][]{drimmel2001} 
seems to be less pronounced that the stellar warp \citep[the Galactic
model; ][]{lopez2002, drimmel2001}, itself less pronounced than the HI
warp \citep{nakanishi2003, levine2006}. However, this conclusion has to be taken
with caution because the error bars are still large and our study does
not go far enough into the external Galaxy.


}

\subsubsection{Dust bar}

{\changes 
The visualisation of our extinction map in the plane Fig.\ref{fig:birdseye} reveals an elongated structure
passing through the Galactic centre. To determine its parameters, 
we fit the straight line to this bar that
minimises the mean absolute deviation of the points of highest density along the bar (Fig.\ref{fig:bar}).

\begin{figure}[t]
  \begin{center}
    \leavevmode
    \epsfig{file=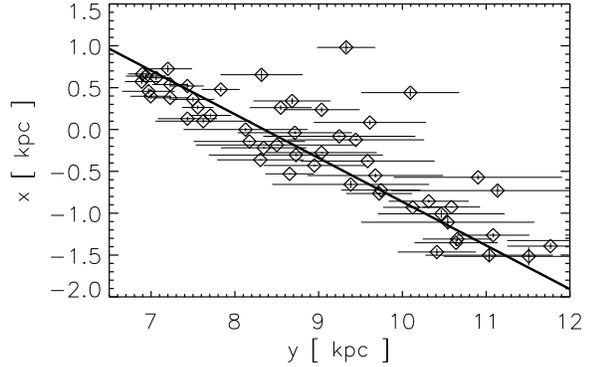,width=\linewidth}
    \caption{Straight line fit that
minimises the mean absolute deviation of the points of highest density along the Galactic bar.}
  \label{fig:bar}
  \end{center}
\end{figure}

We find that the dust bar is at an angle of $\phi=$\barang
relative to the Sun - Galactic centre direction. The structure thus defined has a length of \barlen.
The angle found is higher than the modelled stellar bar (\S\ref{sec:outerbulge}) 
and that found by \cite{babusiaux2005} ($\phi=22.5\pm5.5\degr$), 
but still within the uncertainty of the latter. Furthermore it is compatible with 
the values cited in \cite{gerhard2001}, who find that the angle may lie between $\phi=15\degr$ and $\phi=35\degr$,
 although the bar angles based on stellar studies 
are systematically lower than the value we have found. \cite{bissantz2002} find that COBE/DIRBE observations
 are compatible with an angle for the stellar bar between $\phi=15\degr$ and $\phi=30\degr$, 
but that the model with an angle of $\phi=20\degr$
 fits the observations best.
These results for the stellar bar
 suggest that the dust bar we have detected may in fact be dust lanes which precede the stellar bar at 
negative longitudes and trail it at positive longitudes, 
a feature of our Galaxy suggested by \cite{calbet1996}. 
However, we detect a difference in extinction between positive and negative longitudes of $\sim0.1$ $A_{Ks}$, 
much smaller than their value of between 1 and 2.6 $A_K$.

}

 \begin{figure*}
  \begin{center}
    \leavevmode
    \epsfig{file=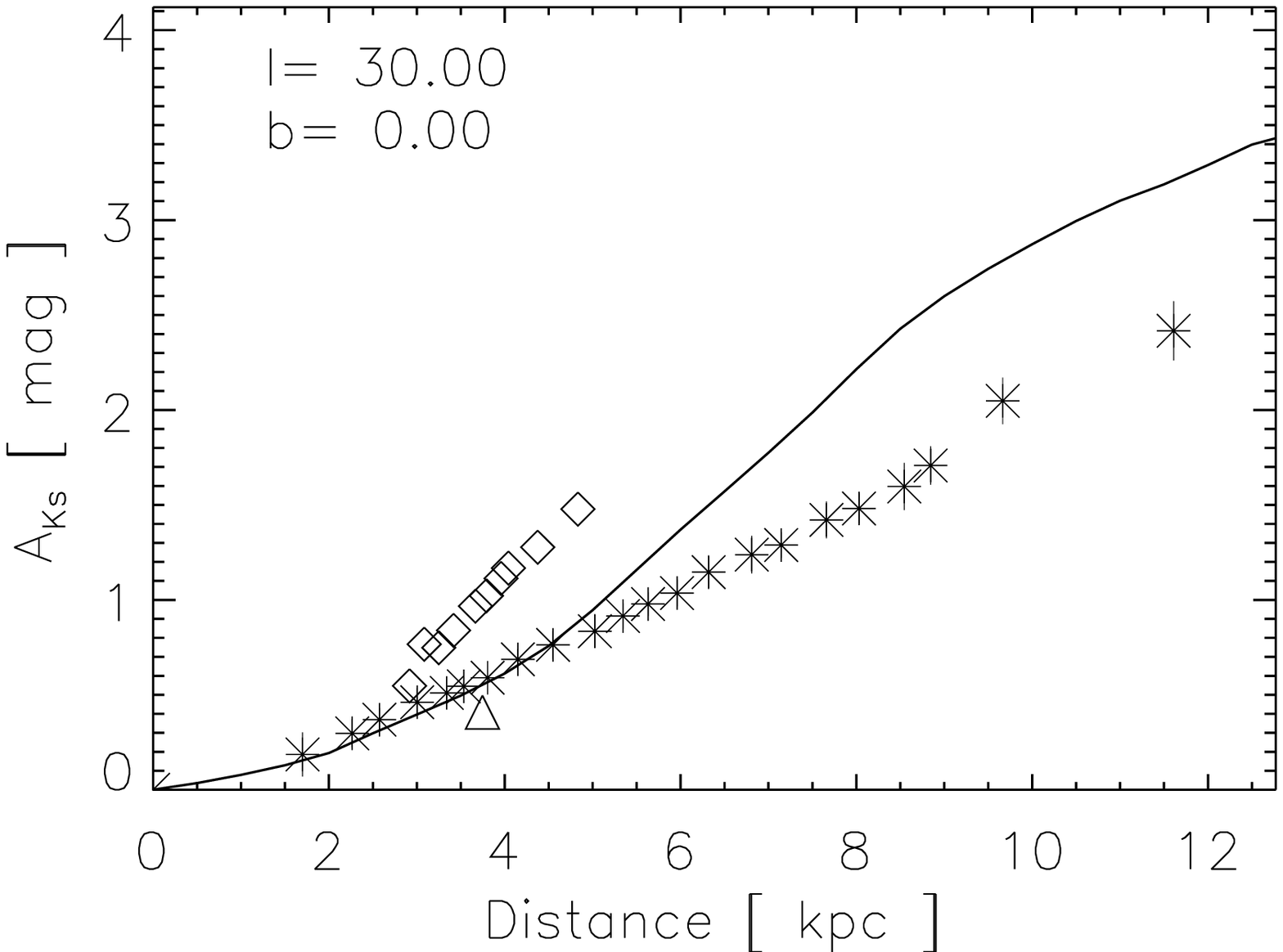,width=0.33\linewidth}
    \epsfig{file=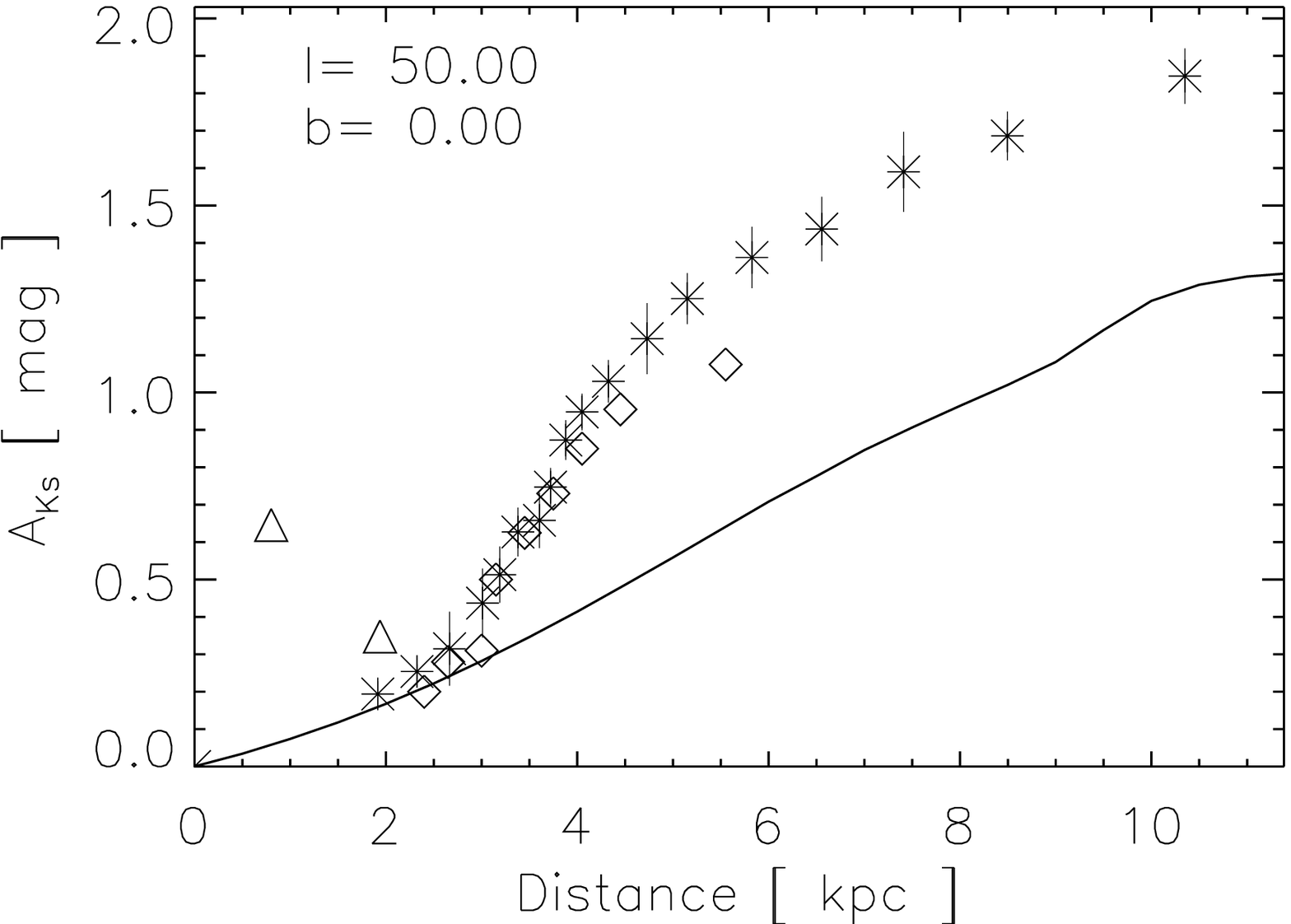,width=0.33\linewidth}
    \epsfig{file=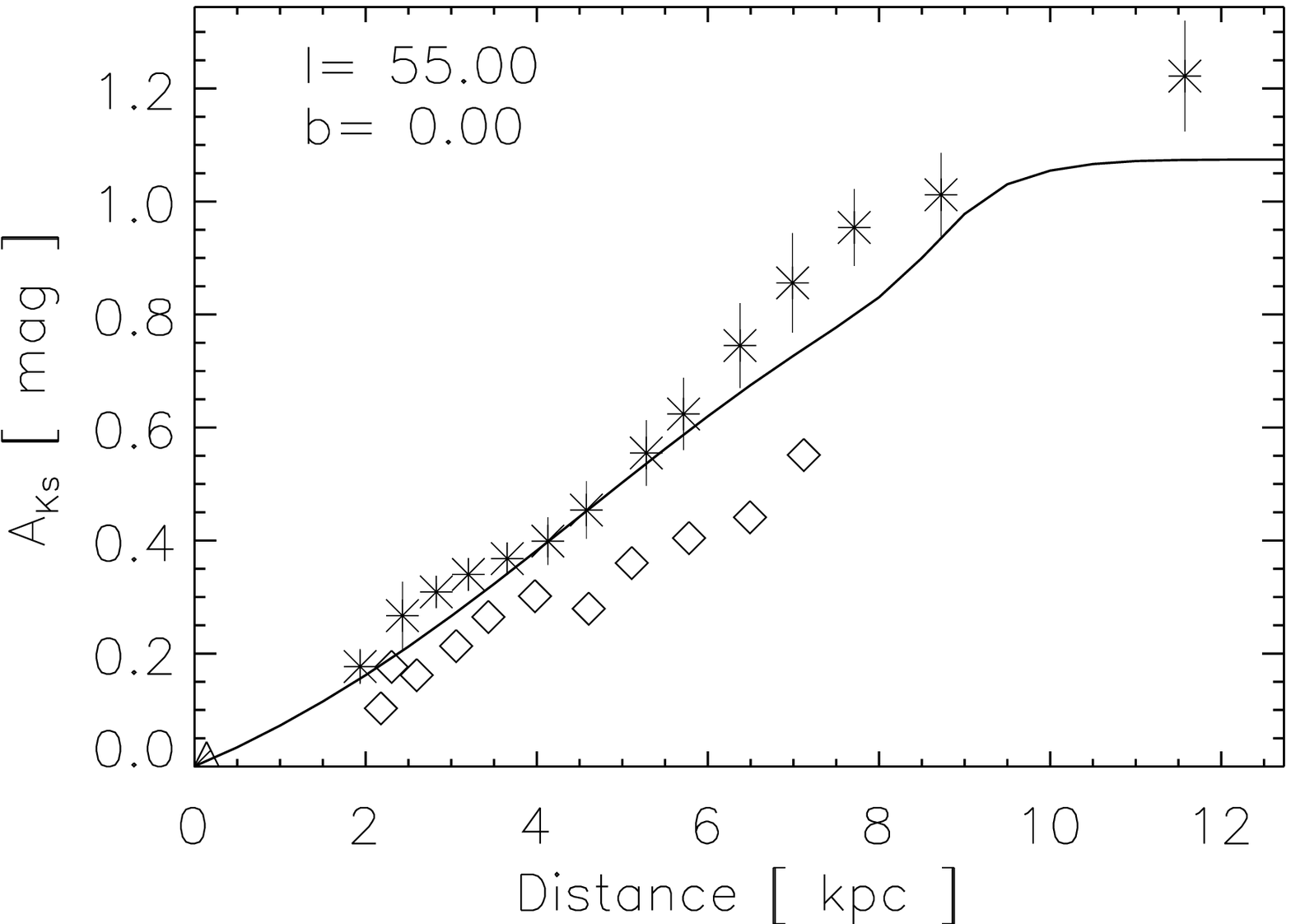,width=0.33\linewidth}
    \epsfig{file=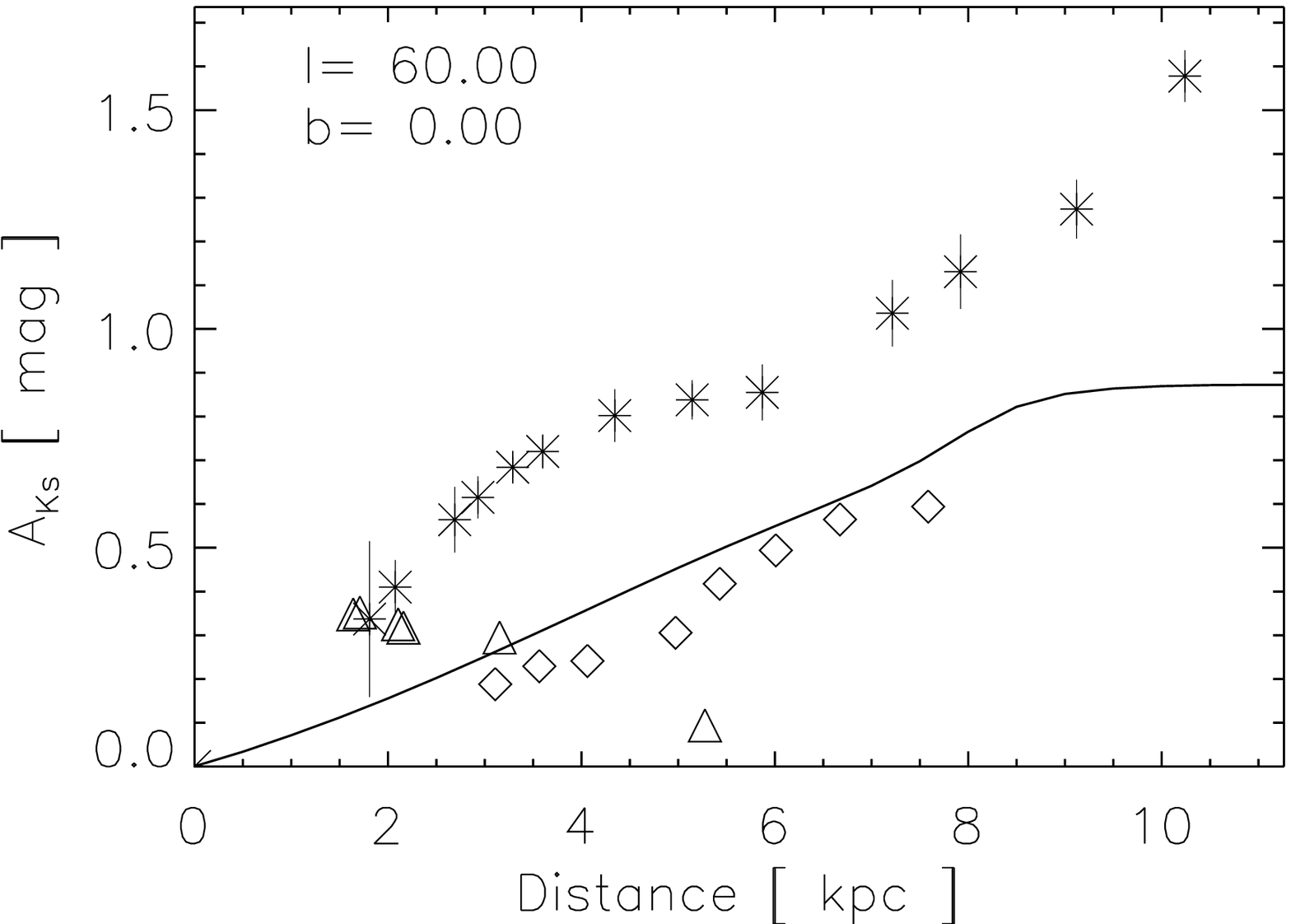,width=0.33\linewidth}
    \epsfig{file=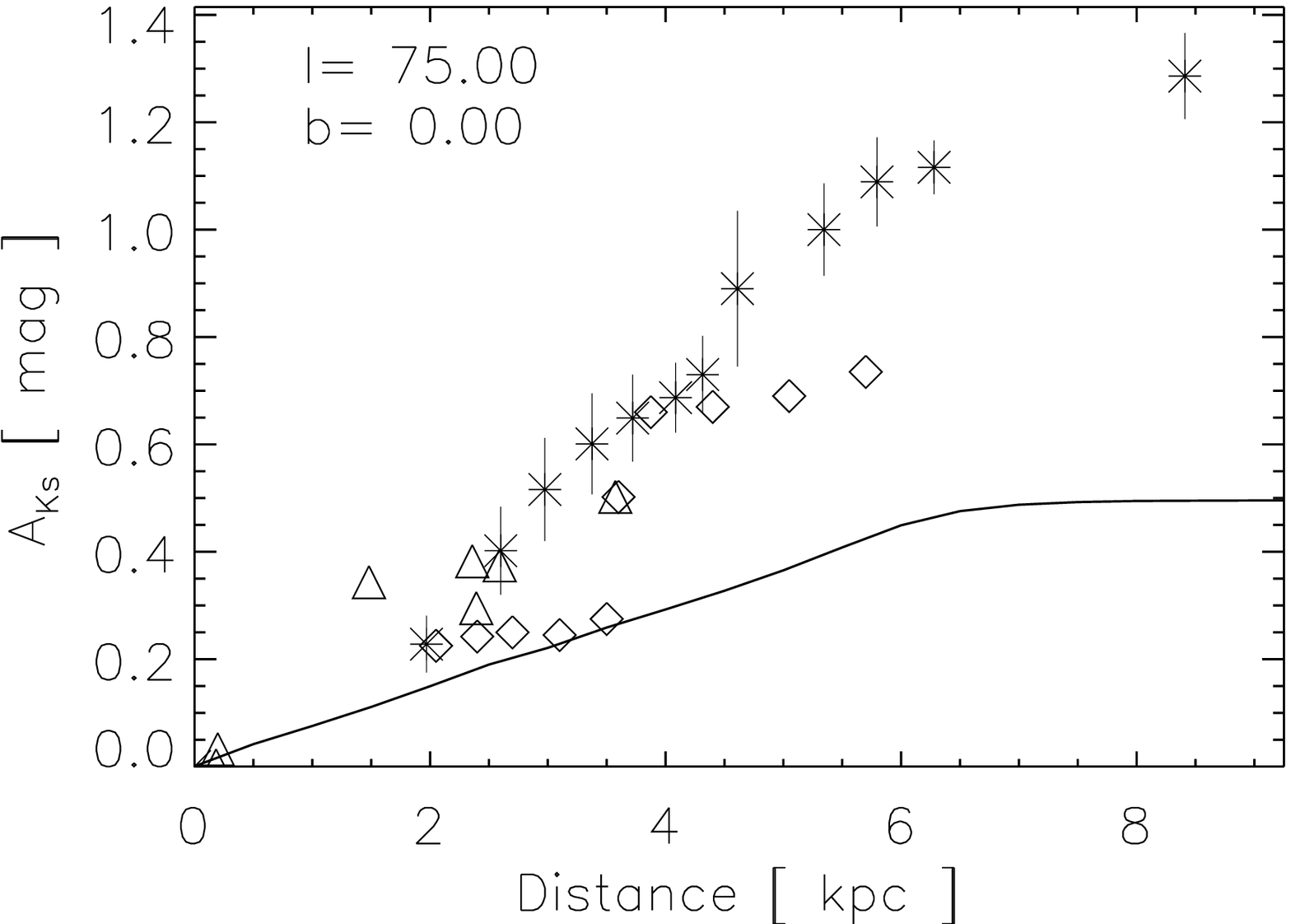,width=0.33\linewidth}
    \epsfig{file=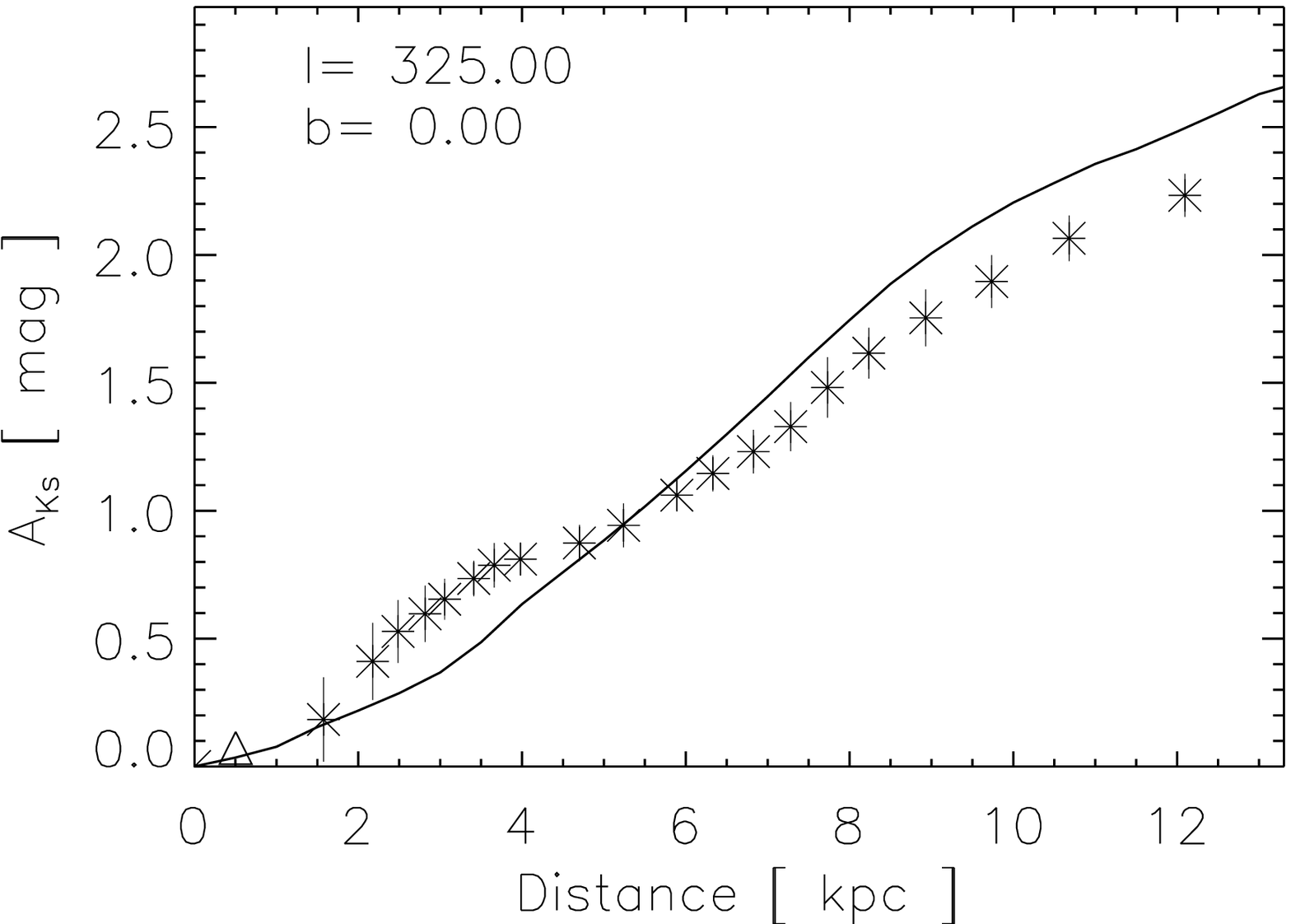,width=0.33\linewidth}
  \caption{The extinction, along the line of sight, in six different fields. For each field, we have
compared our results (asterisks) with the results of \cite{neckel1980} (triangles), 
 \cite{drimmel2003} (solid line) and \cite{lopez2002} (diamonds), where available.} 
  \label{fig:3D_comp}
\end{center}
\end{figure*}
\subsection{Comparison with other 3D methods}

As detailed in \S\ref{sec:intro}, other methods have been developed to measure interstellar
extinction along the line of sight. We now present our results alongside those 
of \cite{neckel1980}, \cite{lopez2002} and \cite{drimmel2003}. 

The three methods are described briefly below; for further information, please refer to the
corresponding article.
\begin{description}
\item[\cite{neckel1980}:] Extinction values and distances are given for over 11000 O-F stars.
The intrinsic colour and absolute magnitude of the stars are deduced 
from the MK spectral type and from 
the H$\beta$ index using the relationship between $M_{V}$ and $\beta$.
\item[\cite{lopez2002}:] This method  
is based on 2MASS observations of red clump giants, for which a value of the 
absolute magnitude and $J{-}K_{s}$ colour is assumed ($M_{K_s} = -1.65$ and $(J{-}{K_s})_0 = 0.75$). 
($J{-}K_s$,$M_{K_s}$) CMDs of the giants are built for 1{\degr} $\times$ 1{\degr} fields and their
displacement from an unreddened CMD, along with their absolute magnitude, are used to calculate 
the extinction and thus
the distance to the extinction.
\item[\cite{drimmel2003}:] The three dimensional dust model by \cite{drimmel2001} is used to calculate
the three dimensional extinction. The dust model is adjusted 
using the far-infrared residuals between the DIRBE 240 $\mu$m data and the 
emission predicted from the model.
\end{description}

Results for various lines of sight are presented in Fig. \ref{fig:3D_comp}.
The \cite{neckel1980} data is the one with the largest dispersion: young and hot OB stars are often
surrounded by their parent cloud and are thus subject to possibly large and varying circumstellar extinction and 
therefore not representative of the whole field.
Our results follow the \cite{lopez2002} results quite closely; this is to be expected as they 
are both based on the near infrared colour difference of giant stars. 
{\changes However significant
differences do exist along certain lines of sight;  this may be due to the fact that they do not use
a model to determine the distance but instead use a unique absolute magnitude for all stars.
Furthermore they do not use the same stellar populations as us, 
isolating the K giants from dwarfs and M giants using the 
``SKY'' model \citep{wainscoat1992}.	In areas where this single population hypothesis 
doesn't hold differences between our results would be expected
}

Finally, we are not always in agreement with \cite{drimmel2003}. This is also 
to be expected given the region under study; anomalous dust temperatures, as well as
 lines of sight passing through spiral arm tangents or near the Galactic 
centre may introduce systematic errors in the \cite{drimmel2003} extinction map. 
Contrary to theirs, our map is not sensitive to dust temperature or to an assumed spiral structure but to the
modelled K\&M giant luminosity function (see \S\ref{sec:lf}).


\section{Discussion}
\label{sec:discussion}

\subsection{Use of the extinction maps}
\label{sec:map_use}
Our results are available in the form of an electronic three dimensional map.
The map is presented in $A_{Ks}$, as the extinction law in this wavelength range is nearly universal
\citep{mathis1990} and so does not depend on the ratio of total to selective extinction.
 By adopting a suitable extinction law (for example \cite{mathis1990}, \cite{rieke1985} or \cite{glass1999}),
one can convert this $K_{s}$ band extinction to an extinction in the visible portion of the spectrum or any
other. 

It was not possible to apply our method to some lines of sight in the region we have studied ($|l|\le$\lmax{\degr},
$|b|\le$\bmax{\degr}) either because there were an insufficient number of stars in the 2MASS observations 
for our pixel size
or the extinction was below the sensitivity of our method (\S\ref{sec:maxmin}). 
In these cases we have substituted the results of our
method with the diffuse extinction disc of the Galactic model (\S\ref{sec:model_ext}), 
with a local normalisation chosen
to minimise the $\chi^2$ test described in \S\ref{sec:chi2}. This solution was also preferred in 
regions where our method could not improve on the extinction predicted by the default extinction of the Galactic model.
{\changes 
To aid the user of our map, we have included details on each line of sight such as the $\chi^2$ statistic and magnitude limits 
of the 2MASS catalogue.  }

{\changes 
Our results are most reliable in fields with homogeneous extinction. As this is not always the 
case, special care should be taken when using fields with  large values for the $\chi^2$ statistic.
In these cases, our method was not able to find a satisfactory solution for the line of sight extinction.
}

\subsection{Effect of changing the model parameters}
\label{sec:change_params}
The determination of extinction using the
Galactic model can be affected by changing the model parameters.
{\changes The magnitude of the extinction that we calculate depends on the difference in colour between observed
and modelled stars; the distance to the modelled stars (and hence to the extinction)
is mostly dependent on the assumed absolute magnitude of the dominant
population.}

This feature of our method means that the number of stars does not affect the extinction determination, 
so that if there are a number of stars in the model which are not present in the observations, we do not
``invent'' extinction in order to remove the excess stars from the model.

 The performance of the Galactic model to reproduce observations in low extinction windows
is a useful measure of the validity of the model. This comparison was done in \cite{robin2003};
star count predictions (stars per magnitude and per square degree) in the $K_{s}$ band were in good agreement  
with observations from the DENIS survey 
and so we refer the interested reader to the aforementioned publication.

We now briefly describe 
the effect on the extinction determination of: the size of the 
hole in the modelled disc distribution (\S\ref{sec:hole});
changing the bulge luminosity function (\S\ref{sec:lf}) ; and 
changing the stellar density \S\ref{sec:scale_length} on the resulting
extinction distribution.

\subsubsection{Galactic disc hole}
\label{sec:hole}
\begin{figure}
  \begin{center}
     \leavevmode
	 {\epsfig{file=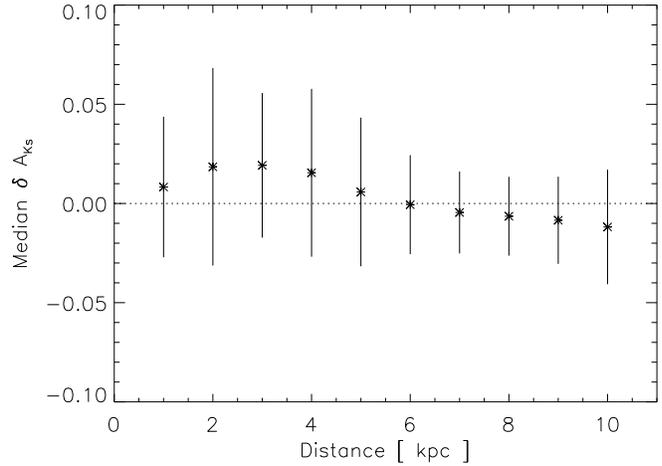,width=\linewidth}}
  \end{center}
  \caption{Influence of the assumed disc density law near 
    the Galactic centre. 
    Median extinction difference, at different heliocentric distances and at $|l| \le 10\degr$ and $|b| \le 2.5\degr$,
    between two determinations of the extinction using models with different values for the disc hole scale length ($R_h$
    in Eq.\ref{eqn:params}). The error bars represent the mean absolute deviation from the median.
    The difference is calculated per pixel, at each heliocentric distance. A positive difference shows that the model with
    the larger $R_h$ predicts a larger value for the extinction. }
  \label{fig:Besac_comp}
 \end{figure}

{
\changes
In Fig.\ref{fig:Besac_comp} we show the median difference between each pixel of 
two extinction maps (map1 minus map2), at different heliocentric
distances and for 
 $|l| \le 10\degr$ and $|b| \le 2.5\degr$.
The Galactic model used for map1 has a central hole in the disc population of 1360 pc
(R$_h$=1360 in Eq. \ref{eqn:params}) and the model used for map2 includes
a hole with a diameter of 680 pc.
The scale in the figure runs from -0.1 to 0.1 $A_{Ks}$; a positive difference indicating more
extinction in the large hole model. As can be seen in Fig.\ref{fig:Besac_comp}, for distances closer
than 6 kpc the large hole model predicts more extinction. We use the Galactic model to calculate the distance
to the extinction:
 with fewer ``distant'' stars in the big hole model, we place the detected extinction closer than for the small
hole model. The median difference is never large, however, and stays within the error of the method(\S\ref{sec:maxmin}).
}

\begin{figure*}
  \begin{center}
    \leavevmode
{\epsfig{file=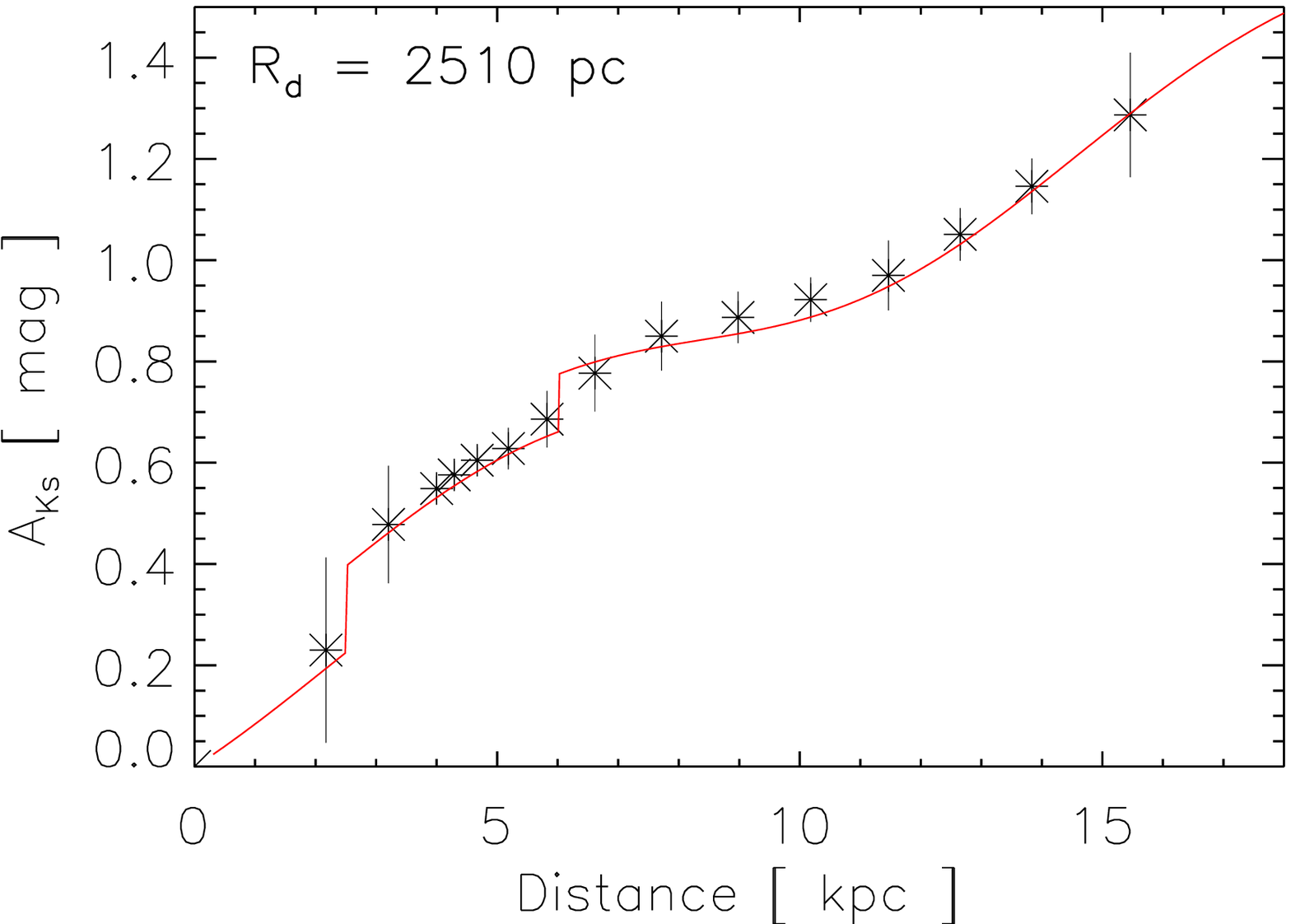,width=0.33\linewidth}} 
{\epsfig{file=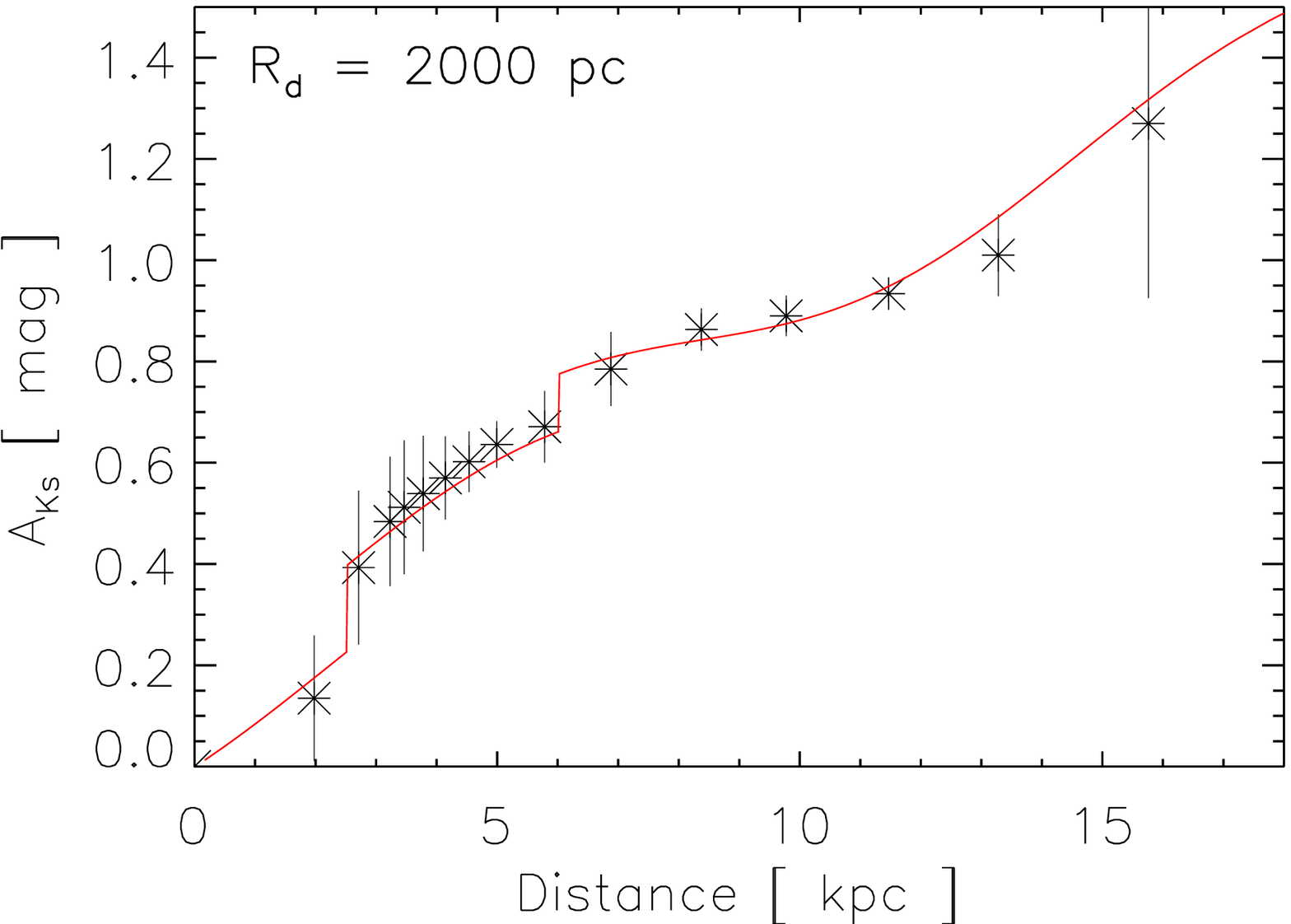,width=0.33\linewidth}} 
{\epsfig{file=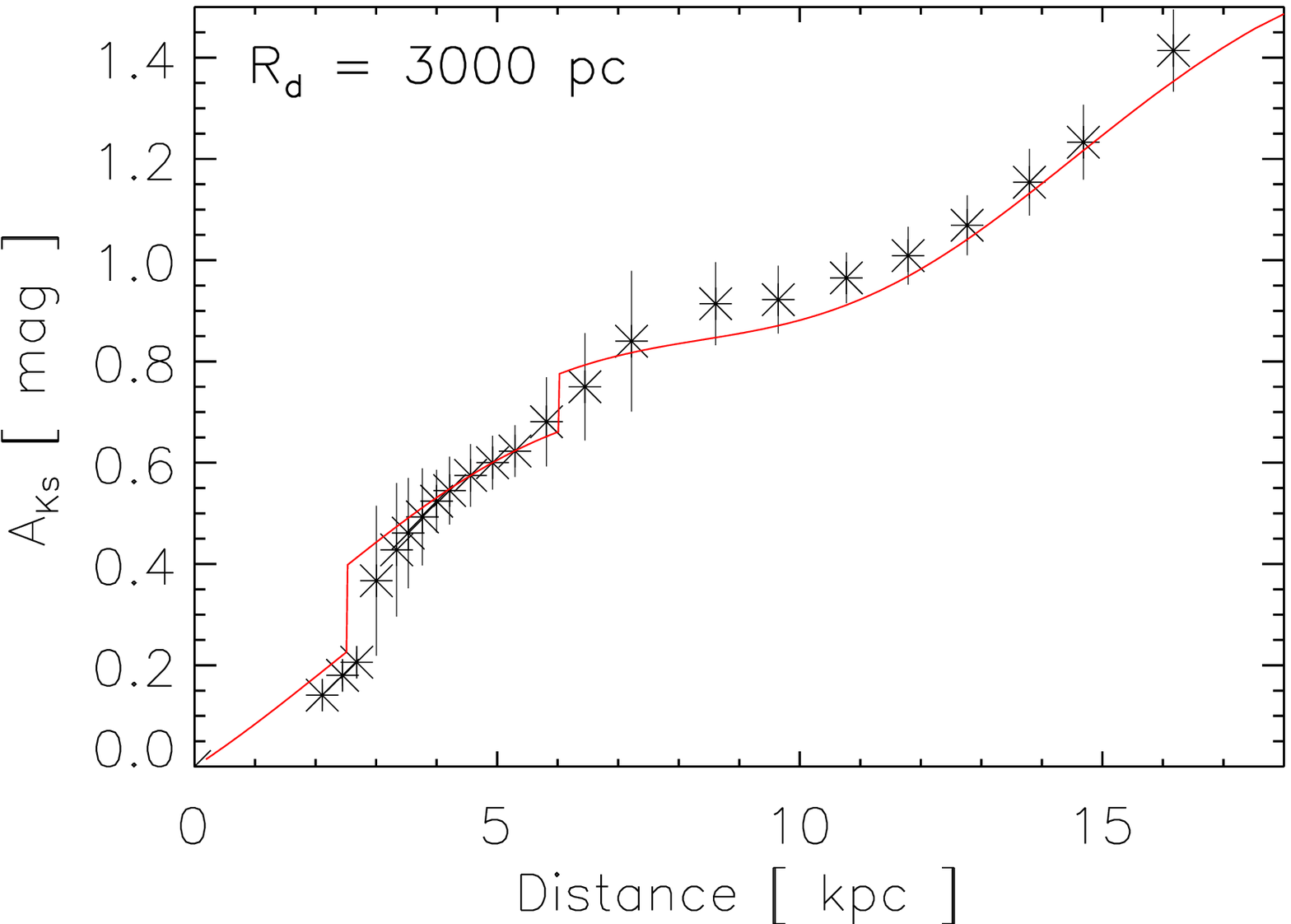,width=0.33\linewidth}} 
{\epsfig{file=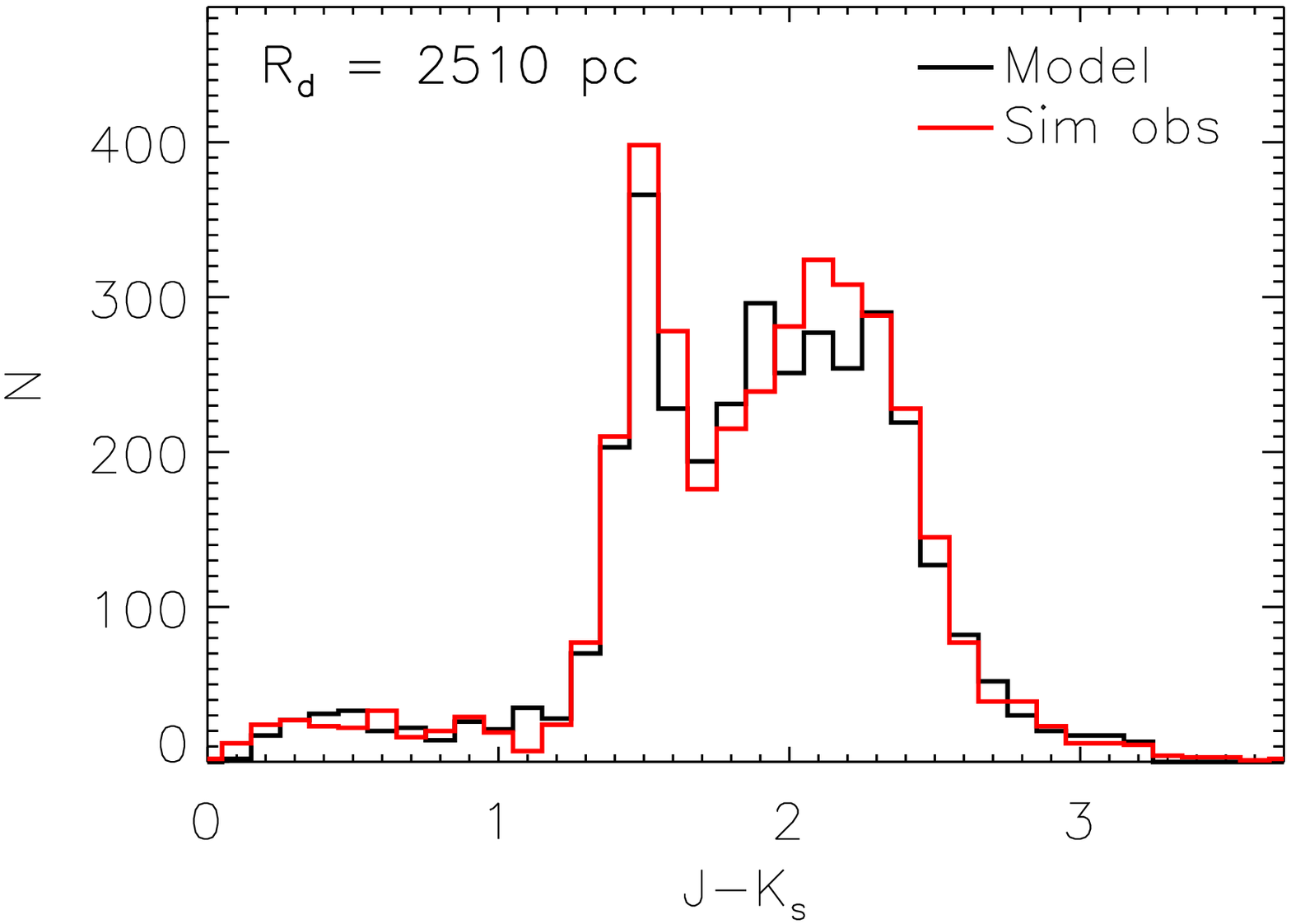,width=0.33\linewidth}} 
{\epsfig{file=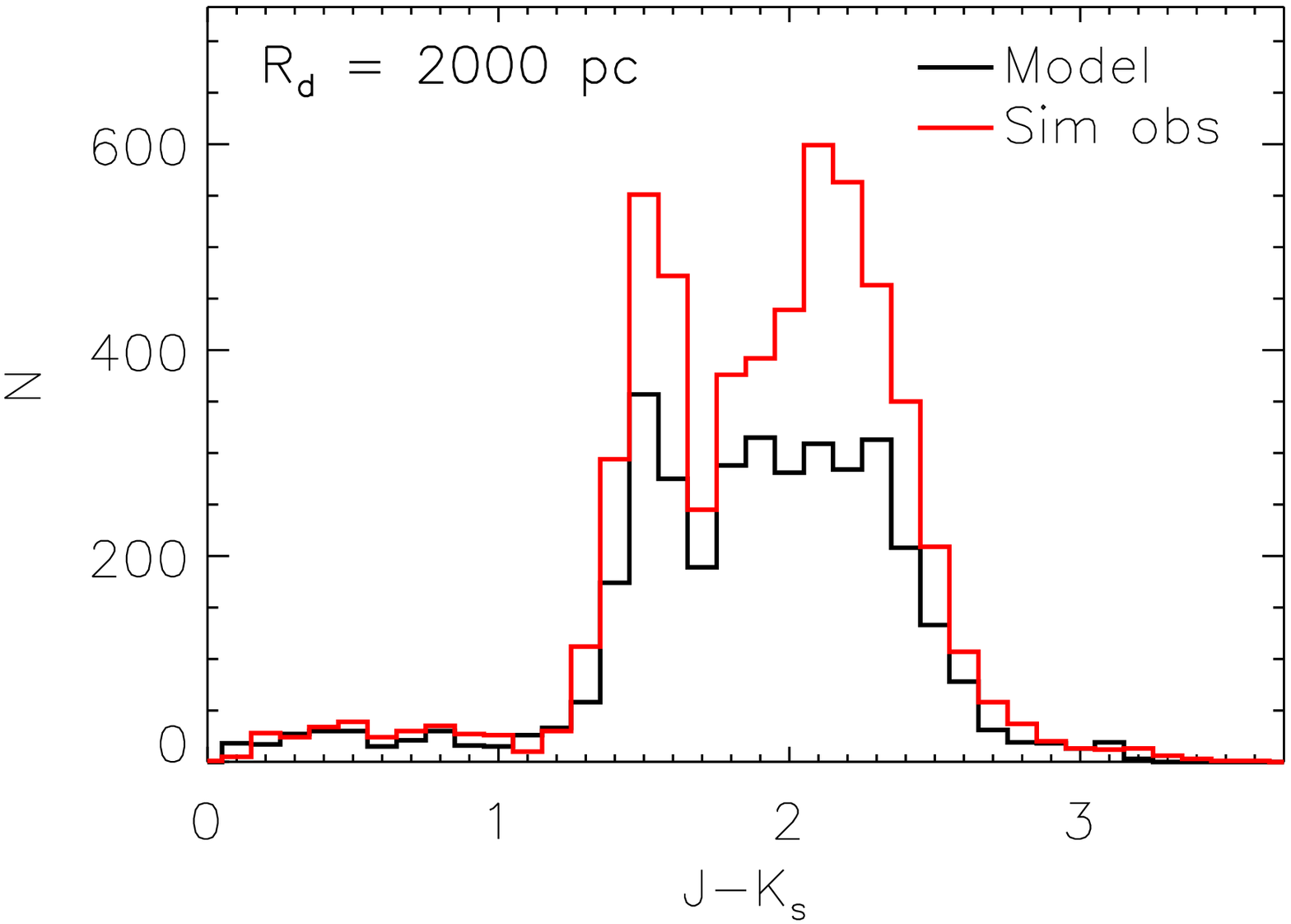,width=0.33\linewidth}} 
{\epsfig{file=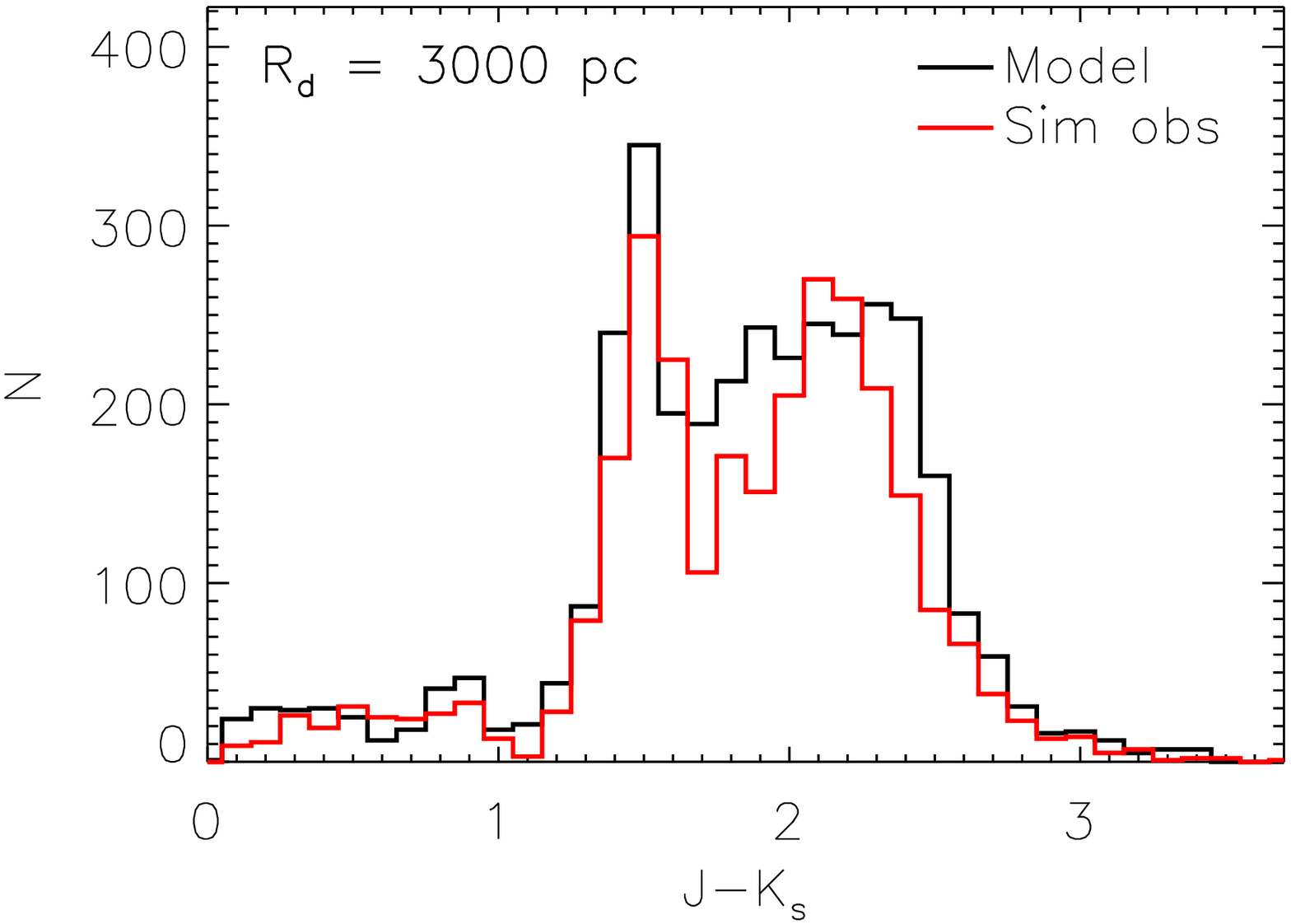,width=0.33\linewidth}} 
  \caption{Influence of the assumed disc scale length. 
Simulated observations are computed with a scale length of 2.51 kpc (first 
column), 2. kpc (2nd col.), and 3. kpc (3rd col.) while the scale length of the model used for 
computing the extinction is assumed to be 2.51 kpc in each case.
{\bf Top}: the distribution of extinction along the line of sight at $(l,b) = (10{\degr},0{\degr})$. 
The solid line represents
    the known extinction law of the simulated observations and the asterisks are the results of our method. In
    each case the scale length used in the simulated observations is included in the graph.
    {\bf Bottom}: the corresponding $J{-}K_{s}$ histograms with the extinction applied.
It is noticeable that the fact that the model 
does not
reproduce the number density of stars in the observations does not 
influence the
determined extinction.}

  \label{fig:scalelength}
   \end{center}
\end{figure*}

\subsubsection{Bulge luminosity function}
\label{sec:lf}
The effect of changing the bulge luminosity function on the determination of the extinction has been
tested. Three luminosity functions were used :

\begin{description}
  \item[{\bf Padova 7.9 \& 10.0 Gyr}]{
    Deduced from theoretical isochrones by the 
    Padova team \citep{girardi2002}, of which we test two bulge ages: 7.9 Gyr, 10 Gyr}
  \item[{\bf Bruzual 10.0 Gyr}] taken from the evolutionary bulge synthesis models 
  of Bruzual \& Charlot \citep{bruzual1997}. We test their 10 Gyr model.
\end{description}
all of them based on a Salpeter initial mass function ($\alpha=2.35$) and assuming a single
epoch of formation (starburst) as well as a mean solar
metallicity (Z $\approx$ 0.02).

Although these luminosity functions alter the distribution of extinction along certain lines of sight,
the overall effect is minimal. The extinction resulting from the use of 
each luminosity function was compared at intervals
of 0.5 kpc for various lines of sight towards the outer Galactic bulge and the maximum difference 
in $A_{Ks}$ between the resulting extinction distributions 
at each distance point was recorded. The ensuing difference distribution 
has a mean of $0.021 A_{Ks}$ and a standard deviation of 0.018 $A_{Ks}$. As this value is comparable to the 
error of our method (\S\ref{sec:maxmin}), we choose to only use the default luminosity function 
for the Galactic model, namely that of Bruzual 10.0 Gyr.

\subsubsection{Galactic scale length}
\label{sec:scale_length}

When modelling the Galactic disc, the density is normalised at the solar position. Therefore,
in the plane of the Galaxy, the stellar density of K \& M giants is largely determined by the 
scale length of the old disc (Eq. \ref{eqn:params}). As the scale length shortens, the number of stars towards the 
Galactic centre increases, while an increase in the scale length has the opposite effect. 

In order to test
the robustness of our method to a change in stellar density, we replace the 2MASS observations in our method with
simulated observations with a known extinction distribution using the Galactic model. 
In the following, we vary the Galactic scale
length in the simulated observations. We then proceed as normal (\S\ref{sec:method}), using
the default model to determine the extinction. This simulates the impact on our method of choosing
a value for the scale length of the Galaxy which is up to 25\% different from the observations.

In Fig.\ref{fig:scalelength} we show the effect of setting the scale length of the simulated observations
to 2510 pc (the default value for the Galactic model), 2000 pc and 3000pc. The top row of this
figure shows the extinction distribution of the simulated observations (solid line) and the extinction distribution
found using our model (asterisks with error bars). 
The simulated observations have been subjected to a diffuse extinction as described in 
\S\ref{sec:model_ext} with a local normalisation of 0.7 mag kpc$^{-1}$ plus
 two clouds at 2 and 6 kpc.
The bottom row shows the $J{-}K_{s}$ histogram of the simulated observations and of the default model with
the calculated extinction distribution.

{\changes The calculated extinction
is found to reproduce the simulated observations' extinction very well, and always within the error bars.}
 At large distances
a small systematic difference is seen :
if we assume a value of 2510 pc for the scale length while setting it to 2000 pc in the simulated observations,
we
produce an underestimation of the 
extinction of $\la$ 10\% at distances larger than 12 kpc. Underestimating the scale length 
(assuming 2510 pc when it is 3000 pc in the observations) 
{\changes results in a slight overestimation of the extinction 
at large distances; at shorter distances, the position of the first cloud is also displaced by 0.5 kpc.
}

\subsection{Bias and uncertainty}
\label{sec:bias}
\subsubsection{Circumstellar dust}
{\changes 

As we are using K\&M giants as a tracer of interstellar extinction, 
our stellar selection will be contaminated by AGB stars with circumstellar dust, making them redder than can
be accounted for by interstellar extinction alone. 
M giants with spectral type later than M5 also show signs of 
circumstellar dust, although the amount has been shown to be negligible \citep{glass2003}.  In most parts of the
Galaxy the ratio of AGB stars to RGB stars  is small, of the order
of a few \%. Only in the inner parts of the Galactic bulge  (at $|l| <$ 2
and $|b| <$ 2) does the fraction of AGB stars become significant (up to $\sim20\%$). However,
number densities of the different types of AGB stars such as
semi-regular variables, Mira-type variables and the high luminosity OH/IR
stars are still very poorly known \citep{habing1996}.

\cite{jura1989} studied the number densities of
mass-losing AGB stars in the solar neighbourhood and estimated their
surface density at around  $\sim 25$ \,kpc$^{-2}$. \cite{lebertre2003}
used the Japanese space experiment IRTS to investigate the
spatial distribution of mass-losing AGB stars.  They found that objects with high
mass-loss in the order of $10^{-6} < M_{\odot}/yr < 10^{-5}$ dominate the
replenishment of the ISM, however these sources constitute a minority of the total number
of stars in our sample ($\sim 10 \%$). 
These stars have a large near-IR excess in $J{-}K$ ($> 0.5\,$mag).

We conclude that there is only a very small level of 
contamination by high-mass losing AGB
stars in our selection, mainly in the region of the inner
 bulge. Our values for the extinction in this region may be slightly overestimated at large distances.
}

\subsubsection{Cloud substructure}
\label{sec:cloud_sub}
For each line of sight measurement, the extinction that we detect is assumed to cover the entire
field at the given resolution. 
However, {\changes \cite{faison2001} have resolved structure 
in the neutral interstellar medium using the Very Long Baseline Array
 at scales as small as 10 AU. 
The resolution of our map is 15\arcmin;  as we cannot resolve structure closer than $\sim$ 1kpc, we will not 
resolve any cloud smaller than $\sim$5 pc across.
}

Nevertheless, as we are placing ourselves in the 
context of the large scale fluctuations in the ISM,
the effects of these small clouds do not have a significant 
effect on our estimation. 
Clouds with an angular size  much smaller than our pixel size will 
introduce anomalous reddening for a small fraction of the stars in a given field.
Our use of the median as an estimator excludes these outlier points.

\subsection{Limits}
 
\subsubsection{Maximum and minimum extinction detectable}
\label{sec:maxmin}

At low $A_{Ks}$, stars that are detected in the $K_{s}$ band are almost certainly detected in the $J$ band.
However, as column density increases and the 
 reddening of the sources gets larger the chances of detecting a source in the $K_{s}$ band
without a corresponding detection in the $J$ band increases.

No selection effect is introduced, however, as we reject the modelled stars in the same way as they are
in the data (same limiting magnitude in the $J$ and $K_{s}$ bands). 
Instead, this will introduce a maximum $A_{Ks}$
in the map as very red stars will not be detected.

This upper limit depends on the magnitude limit in the $J$ band 
of the 2MASS observations.
As we place 
a lower bound on the $K_{s}$ magnitude of 9,
 the highest $J{-}K_{s}$ colour index observable along a particular line of sight would be
$J_{\rm comp}-K_{s {\rm(min)}}$ where the subscript comp signifies the completeness limit and min the brightest
magnitude used due to our self imposed limit. The highest extinction observable along a line of sight
will then be:
\begin{equation}
\label{eq:Ak_max}
A_{Ks(\rm max)}= 0.67 \times \left[ (J_{\rm comp}-K_{s (\rm min )}) - (J{-}K_{s})_{0} \right]
\end{equation}
where ($J{-}K{_{s}})_{0}$ is the intrinsic $J{-}K_{s}$ colour of a K or M giant. 
If we suppose that for a K2 Giant ($J{-}K_{s}$)$_0$ = 0.75 \citep{wainscoat1992},  and knowing that
in the 2MASS observations $J_{\rm comp}$ varies between $m_{J} \sim 12.0$ and 15.8 we
can calculate that the $A_{Ks(\rm max)}$ 
varies between $\sim$ 1.4 and $\sim 3.75$ .
From
equation Eq. \ref{eq:Ak_max}, it
is obvious that the highest extinction detectable will be limited by the completeness in the $J$ band alone.

For a colour difference between 2MASS and the Galactic model to be significant,
the difference must be greater than the uncertainty in the $(J{-}K_{s}$) colour index in the observations.
This 
sets a limit on the minimum extinction detectable.
The photometric errors in the 2MASS observations depend on
the Galactic region observed as well as on 
the local atmospheric conditions on the night the observations were made. An uncertainty of 0.05 magnitudes
for the ($J{-}K_{s}$) colour index is a suitable guide, which corresponds to a minimum $K_{s}$ band extinction
of $\sim$0.03 magnitudes (Eq. \ref{eqn:Ak_basic}). This error becomes 
relatively small for areas of high extinction,
which makes our method ideal for extinction studies in the Galactic Plane.

\subsubsection{Maximum and minimum distances}
\label{sec:maxmin_dist}
In order to allocate a distance to a bin of stars from the 2MASS survey, we make use of the colour
distance relation discussed in \S\ref{sec:colourdist}. To do this we need to exclude the dwarf stars
from our pipeline. As these stars are local, we unfortunately lose most of the information on the first
kiloparsec.

The maximum distance varies according to the $J$ and $K_{s}$ completeness and
column density along the line of sight. Generally, we are able to detect the 
extinction to distances greater than 10 kpc, and for many lines of sight well beyond this, 
but for lines of sight with high column densities and low completeness (central
regions, spiral arm tangents) we may only obtain information out to $\sim$ 7 kpc.
We also notice that lines of sight which contain small
number of stars (large absolute longitudes)  
contain extinction information to rather short 
distances. Deeper counts with more stars would be needed to complete 
those lines of sight.


\section{Conclusion}
\label{sec:conclusion}
We present the first results of our extinction model in the Galactic plane. 
The Galactic model  provides an excellent tool for extracting the reddening information
locked in the 2MASS data.  

Our method does not
give much information on the extinction in the first kiloparsec from the Sun but does give the large scale distribution
of the extinction in our Galaxy. Furthermore, the $J{-}K_{s}$ index is not sensitive to very low
extinction zones, such as high latitude lines of sight. However, the method can easily be modified to use other
colour indices, which would allow us to adapt the sensitivity to a given region or set of observations.
An additional restriction is the number of stars in the observations for our chosen pixel size.
The map could thus be expanded to higher Galactic latitudes by lowering the spatial resolution.

Many structures have been identified in the resulting maps, including: several spiral arm tangents, dust in the 
Galactic bar, the molecular ring as well as local features such as the Aquila Rift.
{\changes We have given quantitative results for three of these structures :
\begin{enumerate}
\item
The dust distribution is found to be asymmetrically warped, in agreement with CO and 
HI observations of the ISM but not as pronounced: for positive longitudes
the angle is \angpos, it starts at \radpos from the Galactic center 
and grows with a slope of \slopepos, while at negative longitudes, the 
angle of the maximum is at \angneg, the starting radius \radneg and the 
slope \slopeneg. Hence we confirm that the warp is seen in all Galactic
components, although the amplitudes appear to vary from one
component to the other.

\item
By converting our extinction map between 4 and 8.5 kpc to a map of hydrogen density, we are able to calculate
the mean scale height for the interstellar matter in the inner Galaxy.
We find the scale height of the disc to be \scheight in the region $4<R<8.5$ kpc. 
This value of the scale height is in agreement with
\cite{drimmel2003} who find a value of $134.4\pm8.5$ pc. However the dust scale height appears to be slightly 
smaller than the HI layer, as seen by  
\cite{malhotra1999} and \cite{nakanishi2003} who find a mean value of $\sim160$ pc for the same region.

\item
The bulge region is found to contain little  absorbing matter apart from  an 
elongated structure \barlen long with an angle of \barang 
relative to the Sun-Galactic centre direction. This dusty bar has a greater inclination than many measures
of the angle for the stellar bar which may indicate the presence of dust lanes preceding the stellar bar at positive longitudes
and trailing it at negative longitudes.
\end{enumerate}
}

In the near future, this 3D map
of the location of obscuring dust in the Milky Way will enable us to constrain dust parameters (temperature, size)
by comparing the predicted far infrared emission with observations.
In addition, this map of the Galactic ISM enables us to further constrain the Galactic 
structure parameters via the Galactic model. 
 In particular, the spiral structure, the Galactic
flare and warp, as well as the size and orientation of the bulge and stellar bar will be easier to model with
this improved estimation of the three dimensional distribution of interstellar extinction.

\section*{Acknowledgments}
We would like to thank the anonymous referee for the very helpful comments we received, which allowed us to improve the final 
version of this paper. Thanks also to Anthony Jones and
 Fran\c{c}ois Boulanger at the Institut d'Astrophysique Spatiale for their comments and suggestions.

This publication makes use of data products from the Two Micron All Sky Survey, which is a joint project of the 
University of Massachusetts and the Infrared Processing and Analysis Center/California Institute of Technology, 
funded by the National Aeronautics and Space Administration and the National Science Foundation. 

{The CDSClient package, available from \url{http://cdsweb.u-strasbg.fr/doc/cdsclient.html},
 was used for the remote querying of the 2MASS dataset.}

M. Schultheis was supported by an APART fellowship.


\bibliographystyle{aa}

\end{document}